# The equilibrium size-frequency distribution of small craters reveals the effects of distal ejecta on lunar landscape morphology

David A. Minton[a], Caleb I. Fassett[b], Masatoshi Hirabayashi[c], Bryan A. Howl[a], James E. Richardson[d].

[a]*Department of Earth, Atmospheric, and Planetary Sciences, Purdue University, West Lafayette, Indiana USA 47907*
[b]*Marshall Space Flight Center, NASA Marshall Space Flight Center, 320 Sparkman Drive NW, Huntsville, AL 35805*
[c]*Department of Aerospace Engineering, Auburn University, Auburn, Alabama USA 36849*
[d]*Planetary Science Institute, 299 E. Lasalle Ave, Apt. 303B, South Bend, IN 46617*

## Abstract

Small craters of the lunar maria are observed to be in a state of equilibrium, in which the rate of production of new craters is, on average, equal to the rate of destruction of old craters. Crater counts of multiple lunar terrains over decades consistently show that the equilibrium cumulative size-frequency distribution (SFD) per unit area of small craters of radius $> r$ is proportional $r^{-2}$, and that the total crater density is a few percent of so-called geometric saturation, which is the maximum theoretical packing density of circular features. While it has long been known that the primary crater destruction mechanism for these small craters is steady diffusive degradation, there are few quantitative constraints on the processes that determine the degradation rate of meter to kilometer scale lunar surface features. Here we combine analytical modeling with a Monte Carlo landscape evolution code known as the Cratered Terrain Evolution Model to place constraints on which processes control the observed equilibrium size-frequency distribution for small craters. We find that the impacts by small distal ejecta fragments, distributed in spatially heterogeneous rays, is the largest contributor to the diffusive degradation which controls the equilibrium SFD of small craters. Other degradation or crater removal mechanisms, such as cookie cutting, ejecta burial, seismic shaking, and micrometeoroid bombardment, likely contribute very little to the diffusive topographic degradation of the lunar maria at the meter scale and larger.

# 1. Introduction

Most of the landscapes of Earth's Moon are dominated by impact craters. The cratered surfaces of the Moon are ideal locations in the solar system for studying the processes by which impact craters shape planetary landscapes. Lunar crater counts that have been calibrated with radiometric dates of samples from associated surface units are a primary tool for dating planetary surfaces across the solar system (e.g. Neukum et al., 2001). If the rate of crater production as a function of crater size is known for a planetary surface, then there should be a direct correlation between the observed number of craters (as a function of crater size) and the exposure age of the surface (Kreiter, 1960; Öpik, 1960; Shoemaker et al., 1963). However, as the cratered surface evolves, the correlation between crater counts and surface age breaks down. Eventually, crater degradation processes cause countable craters to be destroyed at the same average rate that new craters of the same size are produced, and the number of countable craters reaches an equilibrium (Gault, 1970; Shoemaker et al., 1969).

Gault (1970) conducted a foundational experimental study on crater equilibrium, and numerous subsequent studies have addressed different aspects of the equilibrium phenomenon with both observations and modeling (Chapman and McKinnon, 1986; Hartmann, 1984; Hartmann and Gaskell, 1997; Hirabayashi et al., 2017; Richardson, 2009; Woronow, 1985; 1977a; 1977b; Xiao and Werner, 2015). Despite half a century of work on the subject, there remains little understanding of what physical processes determine, in a quantitative way, the observed equilibrium crater size-frequency distribution (SFD) of small lunar craters. Crater equilibrium may be controlled by different processes on different planets and scales. In this paper, we will only focus on one particular type of equilibrium that is seen in populations of small simple craters throughout the lunar maria, which was also the subject of the Gault (1970) experimental study. Our goal is to develop a model that quantifies how the crater production population SFD, and the impact-related processes involved in the formation of each individual crater, contribute to the observed equilibrium SFD. Before we discuss the observational constraints on the type of equilibrium found in populations of small simple craters of the lunar maria, in the next section we will clearly define the basic terms that we use throughout the work.

## 1.1. Definitions of the equilibrium, geometric saturation, and production SFDs

The terms "equilibrium," "saturation," and "saturation equilibrium" have been defined in different ways by different authors (Basaltic Volcanism Study Project, 1981; Gault, 1970; Hartmann, 1984; Melosh, 1989; Woronow, 1977a). In this work, we use the terms "geometric saturation" and "equilibrium" as defined in Melosh (1989) (see also chapter 6 of Melosh, 2011). We use "equilibrium" to refer to the state of a surface at which the formation of a new crater is accompanied, on average, by the obliteration of an old crater. When craters are in an equilibrium state, the production of each new crater is correlated in time with the destruction of (on average) one old one of the same size or larger. It is not necessary for there to be a causal relationship between the formation of one new crater and the destruction of one old crater of the same size. The equilibrium crater distribution was also called the "steady-state distribution" in the Apollo-era literature (e.g. Shoemaker et al., 1969)

Geometric saturation, as defined by Melosh (1989), is a purely mathematical construct that describes the maximum number density of circular features that could be packed onto a two-dimensional surface. The definitions of equilibrium and geometric saturations that we adopt for this work are based on the original terminology used by Gault (1970), with the only difference being that Melosh (1989) added "geometric" as a qualifier to what Gault simply called "saturation." We will quantify each of these definitions shortly, but for context we will first compare them with other terminology used in the literature.

Some authors use the terms "equilibrium" and "saturation" to refer to the specific processes that control the removal or degradation of old craters. Under these process-dependent definitions, "saturation" is used when crater removal is driven predominantly by impact-related phenomena, and "equilibrium" is used when crater removal could potentially include both impact-related instead processes and non-impact related processes, such as wind erosion or volcanic infill. For instance, Chapman and Jones (1977) defined "saturation equilibrium" to mean a situation in which only impact-related processes are involved in the obliteration of craters obliteration, whereas "equilibrium" was a more general term for a surface affected by either impact or non-impact processes.

Hartmann (1984) used the term "saturation equilibrium" in the same way as defined by Chapman and Jones, although Chapter 8 of Basaltic Volcanism Study Project (1981), which was written by a team led by Hartmann, referred to this same phenomenon as "empirical saturation." The same chapter also defines the terms "cookie-cutter saturation," which is the situation when the only crater obliteration process is direct overlap. We note in the case of "cookie-cutter saturation" that the destruction of an old crater is directly caused by the formation of a new crater of the same size, even though in general the condition of equilibrium does not require this causal relationship. The term "complete geometric saturation" also refers to what Melosh (1989) calls "geometric saturation" and Gault (1970) calls simply "saturation."

The various processes involved are not always well constrained or defined, and the terms "equilibrium" and "saturation" are not always defined in a consistent manner. For instance, Gault (1970), and also Marcus (1970), defined "saturation" and "equilibrium" to mean two very different things. In Gault's terminology "equilibrium" is a dynamical phenomenon, in which craters are steadily degraded until they are no longer observable over some finite time, and in that time some number of craters of the same size are generated. It is this competition between the rate of degradation and the rate of formation which sets the equilibrium crater density, regardless of how many circular craters can fit on the surface. "Saturation," in contrast, is defined purely on the basis of geometry, as it quantifies how many craters could potentially "fit" onto a surface

Chapman and Jones (1977) introduced the term "saturation equilibrium" with references to the studies of Gault (1970) and Marcus (1970), which were studying equilibrium. However, Chapman and Jones then define "saturation equilibrium" as the maximum number of craters that can "fit" on a surface before they become destroyed by subsequent craters or covered in ejecta blankets. As Gault defined it, equilibrium was a consequence of a balance between the rates of production and destruction, not as a result of the finite geometry of the surface. While this difference in definitions is somewhat subtle, it is this mixing together of the two distinct ideas of equilibrium with the geometry-based construct of saturation into the amalgamation "saturation equilibrium" which has no doubt been responsible for a great deal of confusion.

The study of Marcus (1970) also did not ascribe equilibrium to geometry, but instead conceptualized it as Gault did as a balance between production and destruction. However, Marcus concluded that the dominant degradation mechanism for small lunar craters was ballistic sedimentation, which he modeled as a diffusive degradation process using the analytical model developed by Soderblom (1970), that we will discuss in more detail in Section 1.3. We note that Marcus (1970) defined "ballistic sedimentation" as both an energetic process that involved diffusive degradation of pre-existing craters and as a low energy "filling-in," or burial, by ejecta.

A further complication in defining and understanding equilibrium is that the way in which equilibrium manifests itself on a given surface depends strongly on the SFD of the crater production population. For the craters relevant to our study, we can model production function as a power law of the form:

$$n_{p,>r} = n_{p,1} X r^{-\eta}, \qquad (1)$$

where, $n_{p,1}$ is a coefficient that gives the cumulative number of craters larger than 1 m in radius per m$^2$ of surface area, and $\eta$ is the slope of the production function. We note briefly that, while it is more common to express crater size as the diameter of the crater rim ($D$), in this work we use crater radius ($r$), as it simplifies the analytical expressions we develop later. In addition, we refer the exponents of power law SFDs as their "slopes," because power law functions appear as straight lines when plotted log-log. Crater SFDs always have a negative exponent, but we define the slope parameter of the production function and other power laws using the positive value for mathematical convenience.

The dimensionless parameter $X(t)$ was introduced in Hirabayashi et al. (2017), and it scales the production SFD by the total accumulated crater density of the surface being studied. We do this in order to remove any explicit dependence on time, because we are primarily interested in how impact-related processes influence the equilibrium SFD and we assume that any impact-related crater degradation processes occur at rate proportional to the crater formation rate. We define $X$ such that $X = 0$ represents the time when the surface under consideration was last initially free of craters, and $X = 1$ represents the present day. We can also write a dimensionless cratering rate as $\frac{d}{dX}(n_{p,>r}) = n_{p,1}r^{-\eta}$.

The equilibrium SFD is also commonly written as a power law, which we define as:

$$n_{eq,>r} = n_{eq,1}r^{-\beta}, \qquad (2)$$

where $n_{eq,1}$ is a coefficient that gives the cumulative number of craters larger than 1 m in radius per m² of surface area, and $\beta$ is the equilibrium slope. In this work, we focus on a particular form of equilibrium that occurs for so-called "steep-sloped" production SFDs, for which $\eta > 2$. For surfaces bombarded by steep-sloped production populations, the equilibrium SFD has a slope $\beta \approx 2$. Figure 1 shows an illustration of how the observed crater counts evolve with time for the case where $\eta > 2$ (similar illustrations appear in Gault, 1970; Melosh, 2011; 1989). At the earlier time, $X_1$, the production function intersects the empirical equilibrium line at $r = 10$ m, and the crater counts (solid line) transition from the production line to the equilibrium line. At the later time, $X_2$, the transition occurs at $r = 100$ m.

While not the subject of this work, we note that when the slope of the crater production function is shallow ($\eta < 2$) the situation is more complicated. It has been suggested using both numerical methods (Woronow, 1978) and analytical methods (Chapman and McKinnon, 1986; Hirabayashi et al., 2017; Richardson, 2009) that the production function slope is preserved in a quasi-equilibrium. In addition, the production cumulative SFD may not have a single slope, and may have segments where $\eta < 2$ over some range of crater sizes and $\eta > 2$ over others. This kind of complex SFD is seen in the lunar production function where the SFD is shallow $\sim 2$ km $< r < \sim 30$ km and then steep for larger craters and basins (Neukum et al., 2001). Multi-sloped production populations may reach a quasi-equilibrium similar to the $\eta < 2$ case that preserves the shape of the production SFD even in equilibrium (Chapman and McKinnon, 1986; Richardson, 2009; Woronow, 1978; 1977a).

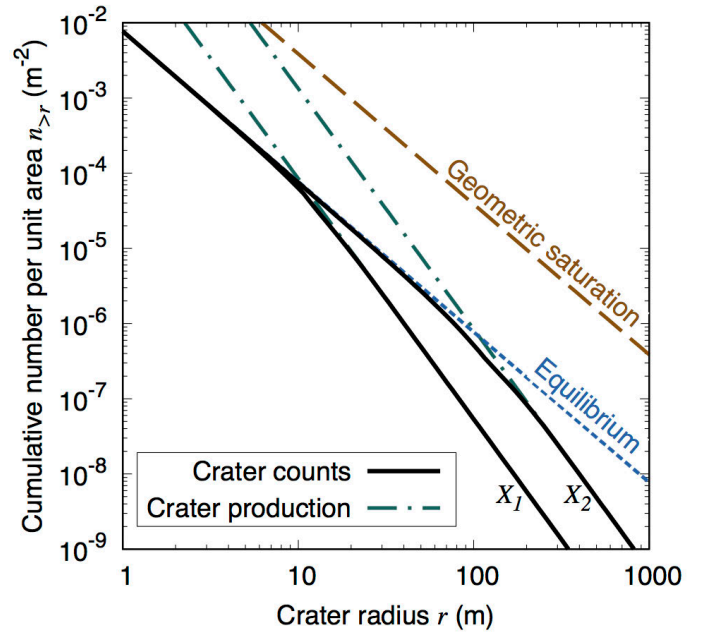

Figure 1. An illustration of the time evolution of the crater count cumulative size-frequency distribution for a steep-sloped ($\eta > 2$) production function. The long-dash line (dark orange) shows the geometric saturation SFD defined by Gault (1970) as $n_{gsat,>r} = 0.385\,r^{-2}$. The short dash line (blue) shows an equilibrium SFD that is $\sim 2\%$ geometric saturation. The dash-dot dark lines (teal) show the production function with a slope of $\eta = 3.2$ at two points in dimensionless time, $X_1$, and $X_2$. The solid black lines show the crater counts at times $X_1$, and $X_2$. At time $X_1$ the production function intersects the equilibrium SFD at $r = 10$ m, and later at $X_2$ the transition occurs at $r = 100$ m (similar illustrations appear in Gault, 1970; Melosh, 2011; 1989).

Equilibrium of the type that develops in steep-sloped crater production functions ($\eta > 2$), as illustrated in Figure 1, were first noted in early imagery of the lunar surface (Moore, 1964; Shoemaker, 1966; Shoemaker et al., 1969) and were the subject of the experimental study of Gault (1970). Because of a combination of the shape of the lunar production function and the age of the lunar maria, the $r \lesssim 100$ m population of maria craters show this steep-sloped equilibrium very clearly. From the Neukum Production Function (NPF), which is a common lunar crater production function summarized

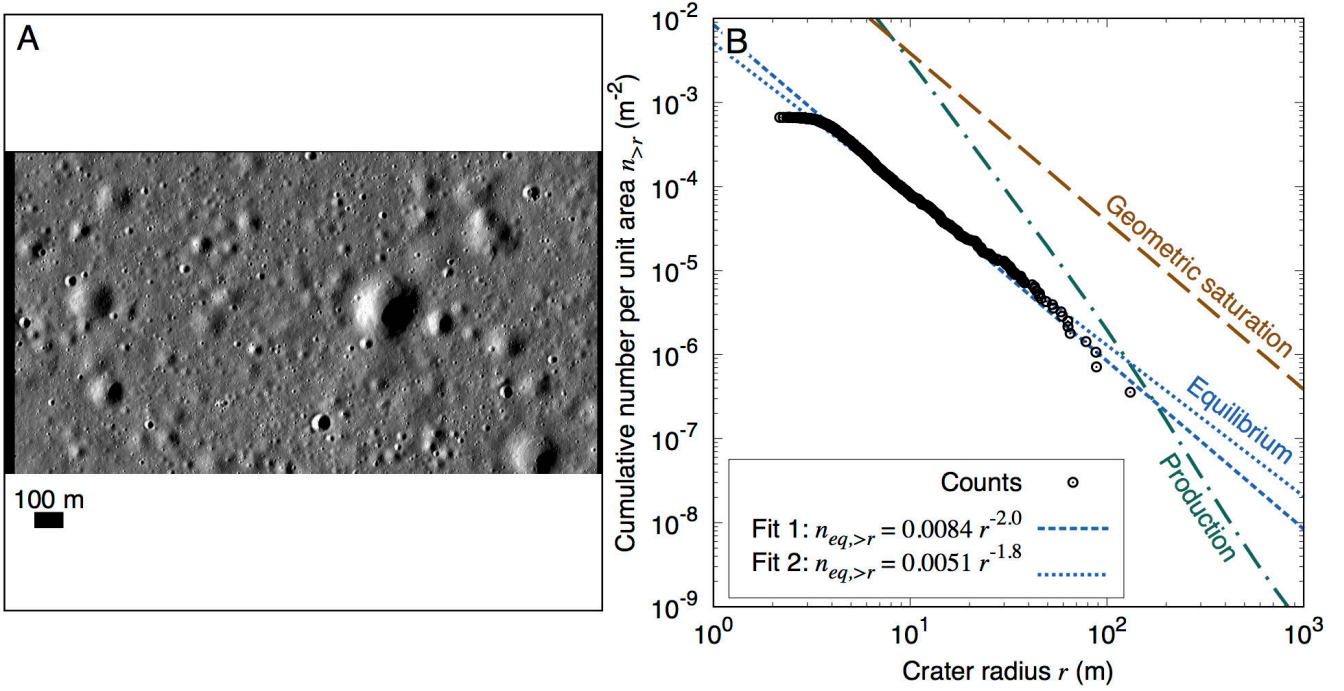

**Apollo 15 landing site**

A

100 m

B

Counts ○
Fit 1: $n_{eq,>r} = 0.0084\ r^{-2.0}$
Fit 2: $n_{eq,>r} = 0.0051\ r^{-1.8}$

Geometric saturation

Equilibrium

Production

Cumulative number per unit area $n_{>r}$ (m$^{-2}$)

Crater radius $r$ (m)

Figure 2. A) A portion of the Lunar Reconnaissance Orbiter Narrow Angle Camera image M146959973L at 0.5 m/pix resolution. This image contains the Apollo 15 landing site in Mare Imbrium was used in the crater count study of Robbins et al. (2014), and is a typical example of a surface in small crater equilibrium. B) The black circles are co-author Caleb Fassett's crater counts of the image in A from the Robbins et al. study. We show two fits for the equilibrium SFD: Fit 1 (blue long-dash) where we constrained $\beta = 2$ and found $n_{eq,1} = 0.0084$, and Fit 2 (blue short-dash) where we allowed $\beta$ to be determined by the fit, and found $\beta = 1.8$ and $n_{eq,1} = 0.0051$ m$^{-0.2}$. The dash-dot line (teal) shows the Neukum Production Function (Neukum et al., 2001) for $n_{D>1\ km} = 5500$ per $10^6$ km$^2$, which corresponds to a model age of 3.54 Gy. The long-dash line (dark orange) shows the geometric saturation SFD

in Neukum et al. (2001), the production function slope is steep ($\eta \approx 3$) for craters with $r \lesssim 2$ km. Nearly all lunar maria formed less than 4.0 Gy ago (Hiesinger et al., 2010), and in that time only $r \lesssim 500$ m craters from the steep-sloped branch of the production function are expected to be above the equilibrium density.

In our notation system the cumulative SFD of geometric saturation is defined to be $n_{gsat,>r} = 0.385 r^{-2}$ (Gault, 1970). The slope value of 2 for the geometric saturation SFD has some important implications that are relevant to understanding observed crater SFDs. Because we define cumulative SFDs as number of craters per unit area, a SFD slope of 2 means that the coefficient of the power law is dimensionless. This is a consequence of what is referred to as "geometric similarity." Objects that exhibit geometric similarity are those that have the same shape at all sizes. Simple lunar craters exhibit a high degree of geometric similarity, and it can be difficult to estimate the scale of images of cratered surfaces with SFD slopes of 2. We will investigate the importance of geometric similarity further as we develop our models later.

### 1.2. Observational constraints on the equilibrium SFD for small lunar craters

Despite the degree of subjectivity that is inherent in crater counting (see for instance Robbins et al., 2014), many different researchers have reached broadly similar conclusions about the equilibrium seen in populations of small lunar craters. Gault (1970) estimated from his experimental results that the equilibrium crater density falls within $1 - 10\%$ of geometric saturation at all crater sizes, which gives an empirical estimate of the line of $n_{eq,1} = 0.021 \pm 0.017$ and $\beta = 2$. Xiao and Werner (2015) counted crater populations on multiple lunar terrains, and found that on terrains where equilibrium occurs at $r < 500$ m, equilibrium was between $0.69 - 3.9\%$ of geometric saturation ($n_{eq,1} = 0.009 \pm 0.006$), which is lower than that estimated by Gault (1970); however, they found that the equilibrium slope was consistently $\beta \sim 2$ across multiple terrains. Hartmann (1984) used observed crater densities across both maria and highlands terrains to construct a similar, though somewhat shallower empirical equilibrium SFD, defined in our notation system as $n_{eq,1} = 0.0064$ m$^{-0.17}$ and $\beta = 1.83$.

An example of equilibrium behavior on a natural lunar surface is shown in Figure 2. The points in Figure 2 shows crater counts of the Apollo 15 landing site counted by co-author Caleb Fassett from the Robbins et al. (2014) crater counter comparison study. In the Robbins et al. study, multiple different human crater counters, both professional and non-professional, were tasked with counting craters on two different images of the lunar surface, and the results were compared to quantify the variability in the resulting data. The Apollo 15 site in Mare Imbrium was one of the terrains studied by Robbins et al. (called the NAC image in the study, as it was $\sim 0.5$ m/pix imagery taken by the Narrow Angle Camera of the Lunar Reconnaissance Orbiter).

We fit two power law functions for the $10$ m $< r < 50$ m craters from Fassett's Apollo 15 crater counts, which are in equilibrium. In the first fit, we constrained $\beta = 2$ (i.e. we impose geometric similarity) and found $n_{eq,1} = 0.0084$, which is 2.2% of geometric saturation. We call this Fit 1 and it is plotted along with the data in Figure 2b. A slightly more accurate fit to the equilibrium SFD is where $\beta = 1.8$ and $n_{eq,1} = 0.0051$ m$^{-0.2}$, which we call Fit 2 and show in Figure 2b. These fits are very similar to each other, and Fit 2 is very close to the empirical equilibrium line defined by Hartmann (1984), using much larger craters on the lunar highlands.

We chose to use co-author Fassett's crater counts of the Apollo 15 site as a major constraint in this work primarily because it is one of the highest quality modern crater counts of an equilibrium SFD available. However, we can demonstrate that our choice of this data as a constraint is well-justified for other reasons. Crater counts are influenced by human factors, and Robbins et al. (2014) found that the overall crater density obtained by different individual crater counters for the Apollo 15 site varied by slightly more than a factor of 2. Co-author Fassett's counts for this site were close to the median value of the ensemble, and so can be considered representative of the site. The equilibrium SFD from our Fit 1 (using Fassett's counts) are 2.2% of geometric saturation, which is well within the range of typical equilibrium crater populations of the lunar maria (Xiao and Werner, 2015).

Robbins et al. (2014) showed that there was much less variation in the slopes of the SFDs between researchers than there was in the overall crater density. In other words, human factors appear to influence $n_{eq,1}$ more strongly than $\beta$. The slope value of $\beta \approx 2$ of our comparison crater counts is also consistent with other slope values obtained by Gault (1970), Hartmann (1984), and Xiao and Werner (2015) on a variety of different lunar surfaces, which suggest that the equilibrium slope of $\beta = 2$ is an important constraint on equilibrium that is not heavily influence by human factors.

The solid line in Figure 2 shows the total production obtained from the NPF for the estimated nominal crater density of the Apollo 15 landing site given by Robbins et al. (2014) of $n_{D>1\,km} = 5500$ per $10^6$ km$^2$ (which corresponds to an absolute age in the Neukum chronology of 3.54 Gy). We fit a power law function to the NPF for $r < 500$ m craters, and find that $n_{p,1} = 4.3$ m$^{1.2}$ and $\eta = 3.2$. From our definition of the dimensionless time parameter, $X = 0$ is equivalent to 3.54 Gy before present in the NPF chronology, and $X = 1$ corresponds to the present-day.

### 1.3. Small crater equilibrium is a consequence of topographic diffusion

The development of small crater equilibrium on the lunar surface is fundamentally a dynamic phenomenon in which the production of new craters and degradation of old craters are in constant competition over billions of years of exposure to an impact flux. In order to understand why the crater count SFD of small lunar craters follows the time evolution shown in Figure 1, and why the equilibrium SFD of found all across lunar surface appears as a power law given by equation (2), with $\beta \approx 2$ and $n_{eq,1}$ a few % of geometric saturation, it is first important to establish a model for how old craters are degraded over time.

Modeling and observations of the morphology of lunar surface features has shown that degradation of lunar landforms is characterized by linear topographic diffusion, or linear diffusive creep (Craddock and Howard, 2000; Fassett and Thomson, 2014; Ross, 1968; Soderblom, 1970). The basic principle is similar to that of soil creep, which has been studied in terms of hillslope in terrestrial environments. (e.g. Culling, 1960; Pelletier, 2008). Landforms undergoing this type of degradation have a characteristically "soft" texture, which was how degraded small craters were described when they were first imaged from spacecraft (e.g. Trask, 1967).

The studies of Ross (1968) and Soderblom (1970) established the framework for understanding one of the ways that the formation of an impact crater can lead to linear topographic diffusion of pre-existing craters. A fresh simple crater is a bowl-shaped depression with a raised rim, and therefore the walls and rim represent local sloped surfaces. Each small crater that subsequently forms on the sloped walls or rim of the larger crater transports some amount of material in its ejecta either upslope or downslope. Due to the mechanics of ejecta transport, more material will be transported at greater distance in the downslope direction relative to the upslope direction. Each small impact that forms inside a pre-existing large crater will induce slope-dependent mass displacement within the large crater, and as long as the length scale of the displacements are small relative to the size of the large crater, then its degradation over time can be modeled as linear topographic diffusion.

In a surface that is undergoing linear topographic diffusion, the evolution of the landscape can be modeled using the linear diffusion equation:

$$\frac{\partial h}{\partial t} = \frac{\partial}{\partial x}\left(\kappa \frac{\partial h}{\partial x}\right) + \frac{\partial}{\partial y}\left(\kappa \frac{\partial h}{\partial y}\right), \quad (3)$$

where $h(x, y)$ is the elevation of the surface at spatial coordinates given by $x$ and $y$, and $\kappa$ is called the topographic diffusivity, which has units $m^2/y$. Diffusivity can be thought of as the "efficiency" or "effectiveness" of the diffusion process. In this form of the diffusion equation, the diffusivity $\kappa$ can vary spatially as $\kappa = f(x, y)$. However, in many situations,

$\kappa$ can be assumed to be a constant over the surface or surface feature of interest, and thus equation (3) may be written in the much more compact form:

$$\frac{\partial h}{\partial t} = \kappa \nabla^2 h, \quad (4)$$

The above linear diffusion model assumes that the diffusivity $\kappa$ does not depend on $h$. Nonlinear diffusion could occur if local slopes are very high (Roering et al., 1999). Advective, rather than diffusive, mass transport is observed in steep walls of relatively fresh lunar craters (e.g. Senthil Kumar et al., 2013; Xiao et al., 2013). However, simple lunar craters only have steep enough walls for these nonlinear processes to be effective for a relatively small amount of their total lifetime, and as shown in Figure 2A, the majority of lunar craters of our comparison data set are very shallow. We therefore assume that nonlinear effects are negligible.

Although the topographic degradation given by equation (4) depends on time, it is more useful for our analysis to remove time in favor of some measure of total number of craters that is independent of any possible time variability in the rate of crater production. If the source of diffusivity is related to the impacts themselves, as in the model of Soderblom (1970), then the rate of diffusive degradation should be proportional the rate of crater production, and therefore the total accumulated number of impacts is a more fundamental parameter instead of elapsed time. We can remove time by defining a quantity called the degradation state, $K$ which is defined as:

$$K = \int_{t_0}^{t} \kappa(t')dt', \quad (5)$$

where $t_0$ is the onset time for a diffusive degradation process of diffusivity $\kappa(t)$. Using this definition, a formal variable change applied to equation (3) gives:

$$\frac{\partial h}{\partial K} = \nabla^2 h. \quad (6)$$

Equation (6) is still a form of diffusion equation, however the explicit dependence on time has been replaced by a dependence on the degradation state, $K$, which has units of $m^2$. Because the change in elevation, $h$, for a given change in degradation state, $K$, depends on the topographic curvature, $\nabla^2 h$, the effect of diffusion on a particular landform depends on that landform's shape. Given an amount of accumulated $K$, features that change elevation over short length scales (i.e. small craters) will appear more degraded than those that change elevation over long length scales (i.e. large craters). The degradation state of a degraded topographic feature can in principle be obtained from measurements of its topography relative to its original, pre-degraded topography (Craddock and Howard, 2000; Fassett and Thomson, 2014).

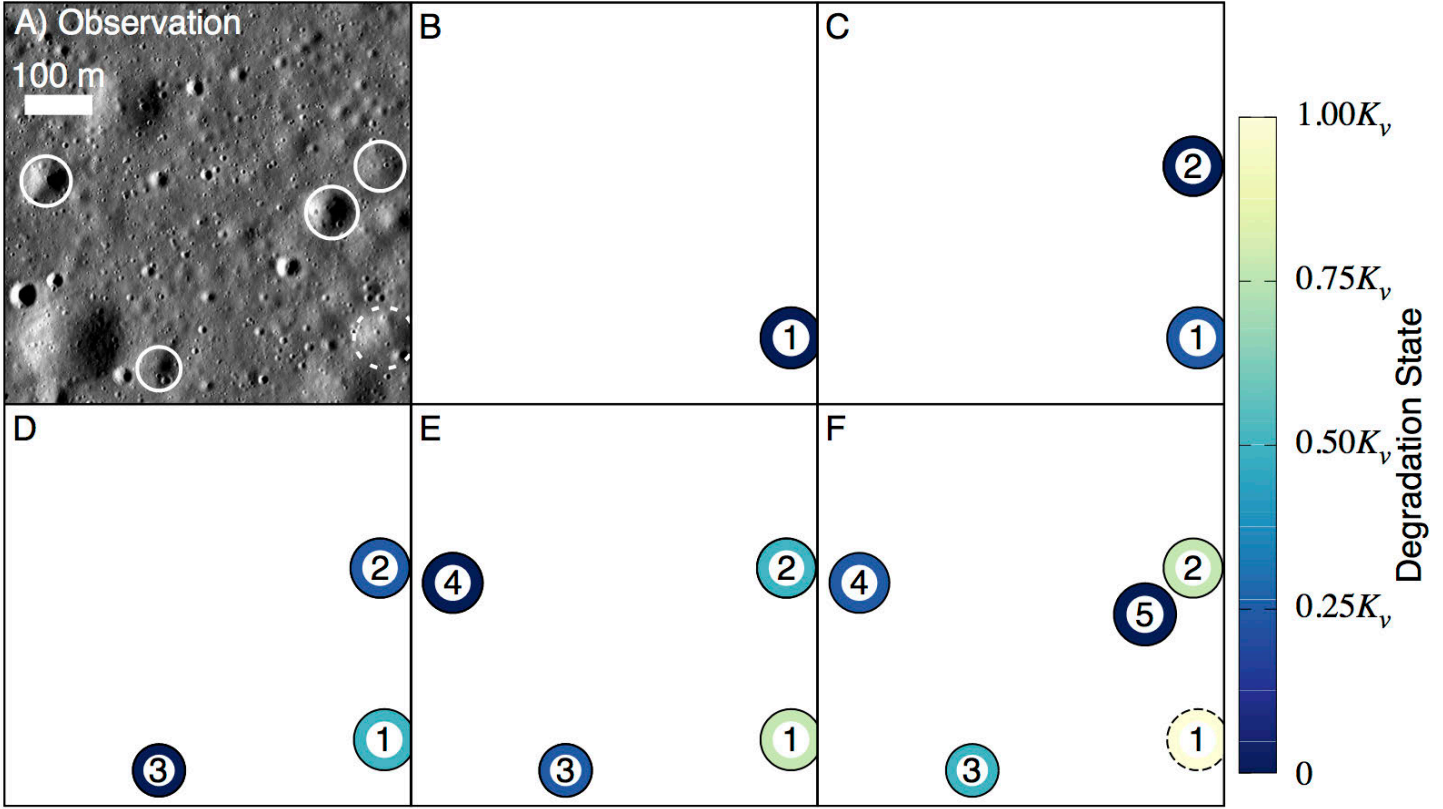

Figure 3. An illustration of the development of crater count equilibrium. Panel A shows a subsection of the image in Figure 2A. The solid circles show the four craters with $30\text{ m} < r < 40\text{ m}$ in this section that were counted by co-author Fassett in Robbins et al. (2014), which are in equilibrium (see Figure 2B). The dashed circle shows a feature that could potentially be a highly degraded crater of the same size, but was not counted. Panels B-F illustrate a simplified model for the cratering history of this terrain. Each panel in the sequence represents the surface after the accumulation of each $30\text{ m} < r < 40\text{ m}$ crater in this size range. The shading represents degradation state, where $K_v$ is the maximum degradation state a crater can have before it is no longer visible. B) The first crater of the sequence forms, and has no accumulated degradation. C-E) As each new crater forms, the pre-existing craters accumulate degradation. E) When the fifth crater forms, the first crater has become too degraded to be counted (dashed circle), and so the total number of craters remains constant, thus the craters are in equilibrium.

The classical diffusion model has been used subsequently to understand how landscapes on airless bodies, such as the Moon (Craddock and Howard, 2000; Fassett and Thomson, 2014) and Mercury (Fassett et al., 2017), evolve due to impact cratering. The importance of diffusive downslope creep induced by small impacts has also been recognized as being related to the observed empirical equilibrium of small craters (Craddock and Howard, 2000; Fassett and

Thomson, 2014; Florenskiy et al., 1977; Richardson, 2009; Richardson et al., 2005; Rosenburg et al., 2015; Ross, 1968; Soderblom, 1970). We demonstrate the connection between (linear) diffusive degradation and the phenomena of crater equilibrium in Figure 3.

Figure 3 shows an illustration of the development of equilibrium on the Apollo 15 site, shown in Figure 2. Panel A shows a subsection of the image of the site. This image shows a terrain with craters of multiple different sizes. Each newly-formed crater is initially a bowl-shaped topographic depression with a raised rim. Subsequent impact-driven diffusive processes, which can be modeled as equation (6), cause the rim of the crater to flatten and the inner bowl to fill in and become shallower as it degrades over time. The craters shown in Figure 3A show a variety of degradation states.

We have highlighted a subset of the counted craters in the figure. The solid circles show the four craters with $30 \, \text{m} < r < 40 \, \text{m}$ in this section that were counted by co-author Fassett in Robbins et al. (2014). Comparing this size range with the crater counts in Figure 2 shows that these four craters are at the equilibrium density. The dashed circle shows a feature that could potentially be a highly degraded crater of the same size but was not counted. Panels B-F illustrate a simplified model for the cratering history of this terrain. In our simplified model a single crater of radii between $30 - 40 \, \text{m}$ forms in each panel in the sequence, and the shading (dark to light) represents the degradation state, $K$, of the crater. In the first model panel (B), the first crater of the sequence forms, and has no accumulated degradation, which we represent by the black shading of the circle. In each subsequent panel (C-F), a single additional crater is added to the surface. Every old crater accumulates some amount of degradation, which we represent with the shading. Lighter shading corresponds to higher degrees of degradation. As craters degrade, they become too degraded to by confidently identified by a human crater counter. This maximum degradation state at which a crater is visible we designate $K_v$. In this example, by the time the fifth crater of the sequence has formed, the first crater has reached a degradation state of $K_v$, and therefore has become too degraded to be counted (dashed circle). At this point the total number of craters remains constant, thus the craters are in equilibrium.

The example shown in Figure 3 also demonstrates why we feel it is important to distinguish between "equilibrium" crater densities and "saturation" (or specifically "geometric saturation") crater densities. From all appearances, the surface shown in Figure 3A contains abundant room for more craters of equivalent size to the four that are circled, and so the surface does not appear to be "saturated" with craters. Instead, the diffusive processes driving crater degradation on this surface operate at a rate such that $30 - 40 \, \text{m}$ radius craters become too degraded to be identified in the time required to generate four craters, and so the total number of craters equilibrates at four.

## 1.4. Crater degradation processes

As we illustrated in Figure 3, the equilibrium crater SFD for small lunar craters is determined primarily by processes that result in diffusive degradation. In Section 2 we will develop an analytical model for equilibrium using linear topographic diffusion, equation (6), as a foundation. Before we develop our model, we will first discuss the processes involved in cratering that lead to degradation of the surface, and a review brief of previous research on modeling the problem of equilibrium cratering and lunar landscape evolution.

The formation of a hypervelocity impact crater is a highly energetic event, and there are a number of resulting processes that contribute to the degradation and obliteration of the pre-existing craters. We illustrate some of the possible mechanisms for degradation in Figure 4A, which includes cookie-cutting, ejecta burial, seismic shaking, secondary craters, preferential downslope deposition of proximal ejecta, and energetic ejecta deposition in distal rays and secondaries. Each of these processes affect the terrain in a different way over different regions around the point of impact, and is worth discussing in more detail.

### 1.4.1. Cookie Cutting

The simplest way that new craters degrade and remove old craters is by cookie cutting. The effect of cookie cutting is to completely render uncountable any old craters whose rims were fully within the rim of the new crater, and so is an important process when a large old craters obliterates all or a portion of any smaller craters that it overlaps (see Fig. 1 of Minton et al., 2015 for an illustration of this process).

### 1.4.2. Low energy ejecta deposition (ejecta burial).

Material excavated from within the crater rim is deposited as ejecta in the surrounding region in the ejecta blanket. Proximal to the crater, within a region bounded by $2 - 3 \times$ the crater radius, the ejecta forms a continuous blanket with a thickness $h$ as a function of radial distance $d$ and crater radius $r$ given by $h = h_{rim}(d/r)^{-3}$, where $h_{rim}$ is the thickness of the ejecta at the rim (McGetchin et al., 1973; Moore et al., 1974; Sharpton, 2014). The exponent of the thickness profile of the proximal ejecta blanket was determined empirically over a large range of crater sizes, from $r = 0.1 \, \text{m}$ laboratory scale craters up to the $r = 593 \, \text{m}$ Meteor Crater in Arizona (McGetchin et al., 1973). Fassett et al. (2011) showed that the ejecta thickness profile of McGetchin et al. (1973) was a reasonable approximation even for the $r = 465$ km Orientale basin, which is one of the largest lunar basins. For $r = 0.1 - 100$ m craters, $h_{rim} = 0.04r$ (McGetchin et al., 1973).

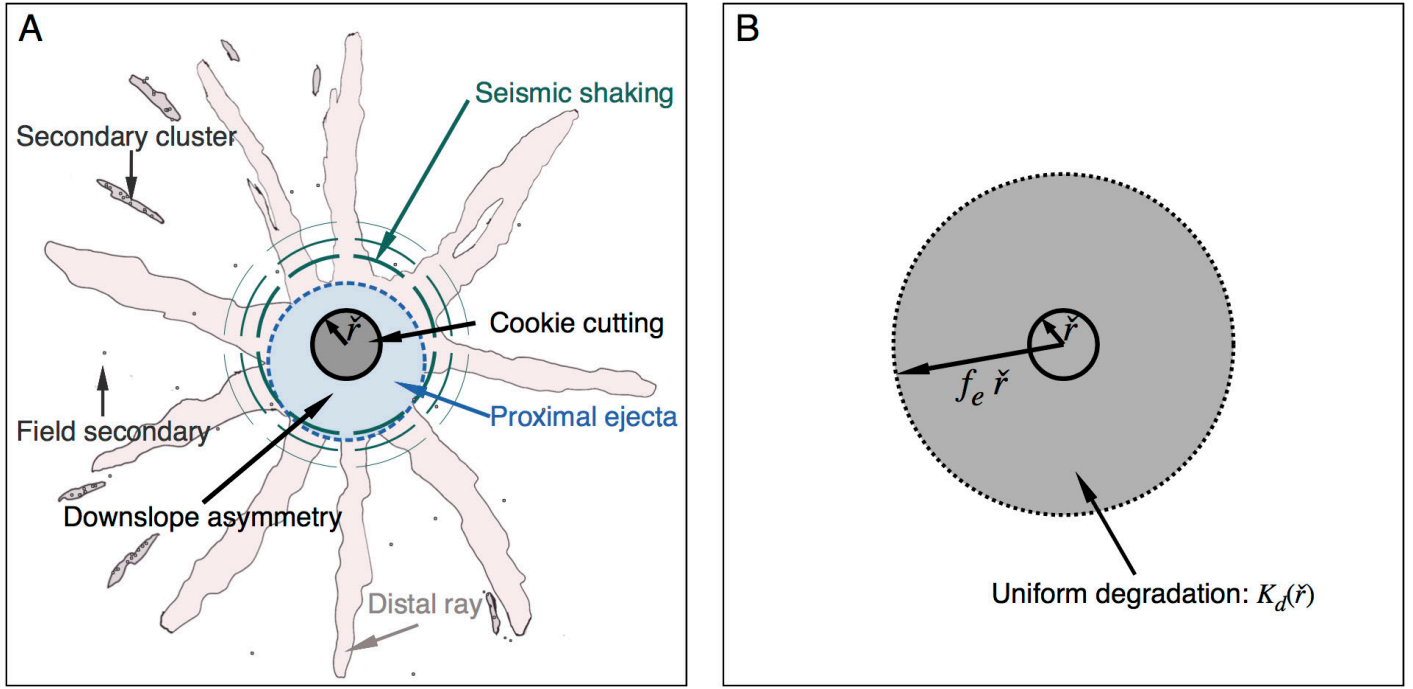

Figure 4. A) A schematic diagram illustrating the spatially heterogeneous nature of impact-driven degradation. In this diagram, several possible degradation processes are shown that accompany the formation of each new crater of size $\check{r}$. Degradation processes illustrated include: cookie cutting within the crater rim (grey), low energy deposition/ejecta burial by the proximal ejecta, which may be asymmetric due to the local slope (blue), high energy deposition of proximal ejecta in rays (light red), field secondaries and secondary clusters (red-grey), and seismic shaking (teal). We approximate the degradation processes shown as a change in the degradation state of the surface by the scalar field $K_c(\check{r}|\rho/\check{r},\phi)$, where $\rho$ and $\phi$ are polar coordinates with respect to the crater center. B) For our analytical model, we model the contribution to degradation given by $K_c(\check{r}|\rho/\check{r},\phi)$ using a uniform, circular degradation region given by $K_d(\check{r})$ over a region with radius $f_e\check{r}$ (grey). The uniform circular degradation contributes the same net average amount of degradation to the surface as $K_c(\check{r}|\rho/\check{r},\phi)$.

### 1.4.3. Diffusive degradation by small impacts (sandblasting).

The formation of proximal ejecta not only buries pre-exiting topography, it also contributes to degradation of the surface by inducing slope-dependent mass transport. Soderblom (1970) developed a simple model based on the idea that the ejecta blanket of a crater that is produced on a slope is asymmetric, with more material preferentially deposited downslope than upslope. This is due to the fact that the ejecta is launched from a conical ejecta curtain that emerges perpendicular to the local surface normal of the impact site. When small craters impact into the rims and walls of large craters, the preferential downslope deposition of their proximal ejecta can induce diffusive degradation of the large craters. This is equivalent to sandblasting (see Fig. 2 of Minton et al., 2015 for an illustration of this process).

### 1.4.4. High energy ejecta deposition (ballistic sedimentation and secondary cratering).

Unlike the relatively spatially uniform blanketing by proximal ejecta, the distal ejecta of an impact crater consists of a spatially heterogeneous population of energetic ejecta fragments that produces crater rays, secondary craters, and induces mass movement and mixing upon deposition in a process called ballistic sedimentation (Elliott et al., 2018; Huang et al., 2017; Oberbeck, 1975).The formation of secondaries in distal ejecta, whether in isolation, in clusters and chains, or as part of distal rays (Elliott et al., 2018; Pieters et al., 1985), should produce similar slope-dependent mass transport of proximal ejecta that the primary proximal ejecta does. Therefore, energetic distal ejecta deposition should lead to diffusive degradation by the same mechanism as that developed by Soderblom (1970). Marcus (1970) identified ballistic sedimentation as the dominant process that degraded the small lunar craters that were the subject of the study by Gault (1970).

### 1.4.5. Seismic shaking.

Richardson et al. (2004) demonstrated that global seismic shaking due to impacts could be responsible for destroying small craters on the Near Earth Asteroid 433 Eros, as the equilibrium slope of craters on Eros is much shallower than the $\beta \sim 2$ slope of standard empirical equilibrium. Additionally, Richardson et al. (2005) showed that the effectiveness of seismic shaking depended on the gravitational acceleration at the surface. They showed that for bodies with diameters larger than 100 km, global seismic shaking became ineffective and seismic shaking could only degrade craters locally for all but the very largest impacts. For the Moon, ($D = 3,474$ km), seismic shaking by impacts is expected to be far less effective at eroding craters than it is on a body the size of 433 Eros ($D = 17$ km), which is supported by the Fassett et al. (2011) results for crater degradation in the region surrounding Orientale basin. Kreslavsky and Head (2012) showed that seismic waves from Orientale formation caused an appreciable degree of degradation on the surface, however an implication of the results of Fassett et al. (2011) is the degradation of craters in the proximal ejecta region of even the largest lunar basins is still dominated by ejecta burial over seismic shaking. Therefore, seismic shaking is unlikely to be an important process for setting the equilibrium SFD of lunar maria craters.

### 1.5. Previous work on modeling equilibrium

A successful model for equilibrium should quantify how much each of the various crater degradation and removal processes discussed in Section 1.4 contributes to setting the equilibrium SFD. Ross (1968) developed an analytical model for crater degradation based on impacts by smaller craters (see Section 1.4.3). Ross was modeling the general problem of crater degradation, not the equilibrium problem specifically. He used the fact that the distribution of ejecta from craters on sloped surfaces is asymmetric, with more mass deposited on the downslope side than the upslope. He showed that this slope-dependent mass asymmetry of small crater ejecta was an important process for the degradation of $r < 500$ m lunar craters.

Soderblom (1970) later showed that the slope-dependent mass asymmetry of Ross (1968) could be modeled with linear diffusion (see Section 1.3), and investigated the problem of equilibrium. Soderblom considered secondary craters in his model, but concluded that a model without secondaries fit the observed equilibrium SFD (which he termed the "steady-state" distribution) better than one with secondaries. Marcus (1970) also investigated the same problem as Soderblom with a similar analytical model, and came to a somewhat different conclusion. He noted that degraded small craters have much flatter floors than fresh craters, and on this basis determined that low energy ejecta deposition (see Section 1.4.2), which he termed a form of ballistic sedimentation, was important in determining the equilibrium SFD. Though the analytical models of both Soderblom (1970) and Marcus (1970) were sophisticated, and formed the basis of a later model by Hirabayashi et al. (2017) as well the one we develop here, both models contained assumptions that make it hard to use them to set quantitative constraints on the role of secondaries and low energy ejecta deposition (see Hirabayashi et al. 2017).

The analytical models Soderblom (1970), Marcus (1970), Hirabayashi et al. (2017), and the models we will develop in Section 2 of this paper all represent the population of observable craters as a 1-D SFD that evolves in time due to the production of new craters and the degradation and removal of old craters. These kinds of models make a simplifying assumption that the 3-D topography of the individual craters can be approximated as a single parameter that characterizes a crater's visibility to a human crater counter. These models also assume that SFD represents an average of the 2-D spatial distribution of craters on the landscape.

The study of Gault (1970) is a unique experimental study of crater equilibrium. In this study, Gault produced craters by firing projectiles of various sizes into a 2.5 m × 2.5 m sandbox at NASA Ames Research Center. He studied several different production SFDs, from shallow to steep. Based on his experimental results, he concluded that in steep-sloped

SFDs, diffusive degradation by the numerous small projectiles in combination with ballistic sedimentation was the dominant crater degradation mode, though these conclusions were somewhat qualitative.

Many studies of crater equilibrium have employed numerical models that represent the 2-D distribution of craters on the landscape. Woronow pioneered the use of numerical methods to study equilibrium cratering in late 1970s and early 1980s, using both Monte Carlo and Markov Chain methods to model craters on 2-D surfaces (Woronow, 1985; 1978; 1977b; 1977a). Woronow's models represented as circular features corresponding to the crater's rim, and modeled how the formation of new craters could remove or degrade the rims of pre-existing craters. Similar 2-D Monte Carlo "circular rim" codes were also developed by both Chapman and McKinnon (1986) and Marchi et al. (2012). Most of these studies were focused on the problem of understanding equilibrium in large lunar craters and basins, where the production function is not a simple power law and contains shallow ($\eta < 2$) branches. However, Marchi et al. (2012) performed simulations of a lunar maria case where $\eta > 2$ as a way of calibrating the code's ability to reproduce the equilibrium SFD.

Unfortunately, due to the nature of these 2-D circular rim codes, it is difficult to relate the results to the physical processes involved in the diffusive degradation of lunar craters. For instance, the primary free parameter in the code of Marchi et al. for generating the observed equilibrium SFD of small lunar craters is a factor that determines whether or not a crater of a given size can destroy the rim of a crater of a larger size. That is, new craters of size $f\check{r} > r$ can remove the rim of an old crater of radius $r$. They found that $f = 9$ gave a match to the equilibrium crater counts of the Sinus Medii mare by Gault (1970), but it is not clear how this factor is related to physical processes involved in crater degradation, such as those discussed above and illustrated in Figure 4A.

As shown in Figure 3A, small lunar craters are degraded by steady diffusive degradation that makes them shallower over time. In this degradation mode, the entire crater is affected by degradation and it is likely that the visibility of a crater involves both the rim and inner bowl (Fassett and Thomson, 2014; Ross, 1968; Soderblom, 1970). Therefore, 2-D circular rim codes may have limited ability to model the processes involved in setting the small crater equilibrium SFD. To model the diffusive degradation of craters in the steep-sloped production function regime ($\eta > 2$) regime, a 3-D landscape evolution code that models the evolution of the circular depressions that defines craters on a landscape has a strong advantage over a 2-D circular rim code.

The first 3-D landscape evolution code used to study crater equilibrium was the GASKELL code, which was used in Hartmann and Gaskell (1997). GASKELL generates three-dimensional digital elevation models (DEM) of the simulated landscape. In Hartmann and Gaskell (1997), a Monte Carlo cratering model was added to GASKELL that produced realistic 3-D morphology of craters and their ejecta blankets on a DEM that simulated a heavily cratered surface of Mars. For the Hartmann and Gasskell study, the DEMs of the simulated cratered landscapes were converted to simulated imagery and the simulated craters were counted by human crater counters.

Hartmann and Gaskell (1997) performed several simulations using steep-sloped production SFDs. In some of their simulations they produced terrains with countable crater SFDs that significantly exceed the observed equilibrium SFD, which is contradictory to observations of natural surfaces. They proposed that this mismatch between simulation results and observations could be solved if they were to include the collective effects of small secondary craters that were below the resolution limit of the simulation. By including an ad-hoc model for this effect, they were able to create crater SFDs that matched the observed equilibrium SFD. However, Hartmann and Gaskell (1997) did not quantify this sub-pixel degradation model. The study of Hartmann and Gaskell was specifically about the cratered landscapes of Mars, where fluvial erosion likely dominated the degradation of the ancient heavily cratered terrains (Craddock et al., 2018). However, the study of Hartmann and Gaskell (1997) should, in principle, be generalizable to surfaces where fluvial effects are not important, such as the small lunar maria craters.

Another 3-D landscape evolution model used to study equilibrium is the Cratered Terrain Evolution Model (CTEM), which was used to study the question of equilibrium of the large craters of the lunar highlands in Richardson (2009). The CTEM code is similar to the GASKELL landscape evolution model of Hartmann and Gaskell (1997), but with a number of important differences. Like GASKELL, CTEM produces DEMs of simulated heavily cratered surfaces, but unlike GASKELL the craters are counted automatically by the code, rather than manually. CTEM also includes ejecta burial and seismic shaking models based on linear topographic diffusion.

Richardson (2009) studied crater equilibrium on the large craters and basins of the lunar highlands and reached broadly similar results as Chapman and McKinnon (1986), which employed a 2-D circular code. Both studies concluded that the heavily cratered lunar highlands could be in an equilibrium condition and that the complex shape of the crater count SFD of the lunar highlands could be attributed to the complex shape of the production population.

However, both the design of the simulations and the state of the CTEM code at the time of Richardson (2009) make it difficult to relate its results to the underlying processes involved in setting the equilibrium SFD. For instance, the automated crater counting algorithm used in Richardson (2009) was not calibrated by a human crater counter. The early version of CTEM also lacked a constrained model for the effects of sub-pixel cratering, just as the GASKELL code did. In addition, the production SFD was artificially truncated in the Richardson (2009) study such that no crater with a diameter larger than 50% of the domain size was produced. A number of improvements to CTEM were implemented in a later study by Minton et al. (2015). Minton et al. showed that the removal of the artificial truncation of the production SFD for the lunar highlands lead to very different evolution of the lunar highlands compared to what was shown in Richardson (2009). However, the study of Minton et al. (2015) did not directly address the problem of crater equilibrium.

Observational studies of the topographic evolution of the lunar landscape also have relevance to the small crater equilibrium problem. Fassett and Thomson (2014) used observations of the topographic profiles of simple lunar craters in the size range $400 \, \text{m} < r < 2500 \, \text{m}$ to estimate the degradation rate of the lunar surface. They placed constraints on the value of topographic diffusivity, given by the $\kappa$ parameter in equation (4), as a function of time for craters in their observational size range, but did not explicitly model the small crater equilibrium SFD nor did they constrain the processes involved in determining $\kappa$. Craddock and Howard (2000) developed a similar model for linear topographic diffusion and applied it to the degradation of lunar craters in the size range of $500 \, \text{m} < r < 1500 \, \text{m}$. In that work, they assumed that the dominant process by which craters in this size range are degraded is micrometeoroid bombardment (e.g. the flux of $\sim 1 \, \text{mm}$ primary impactors).

In the following sections we will develop diffusion-based models of small crater equilibrium. Our approach combines an analytical 1-D model for the evolution of the observable crater SFD with a 3-D Monte Carlo landscape evolution code. We first develop our analytical models in Section 2. Our analytical models are similar to the a 1-D models of Soderblom (1970), Marcus (1970), and Hirabayashi et al. (2017) in which we represent the observable craters as an time-evolving SFD. In Section 3 we will model the landscape using the CTEM code used in the study of the lunar highlands equilibrium in Richardson (2009), with added modifications introduced by Minton et al. (2015), Huang et al. (2017), and in this work. Because CTEM represents the full 2-D spatial distribution of craters on the landscape, as well as the 3-D morphology of each individual craters, we will use it to test the robustness of the assumptions inherent in the 1-D model analytical developed in Section 2. We will model several of the important processes in the degradation and removal of craters, including as cookie-cutting, low energy ejecta deposition (e.g. ejecta burial), high energy ejecta deposition (e.g. ballistic sedimentation and cratering by small secondaries), and bombardment by primary micrometeoroids. Our goal is to investigate which of the proposed impact-related processes are important in determining the equilibrium SFD of small craters of the lunar maria.

## 2. A linear diffusion model for small crater equilibrium

In this section we develop a diffusion-based model for small crater equilibrium. In this model, we track the SFD of the observable crater population as the craters are accumulated and degraded on the surface. The underlying mathematical concept of our model is very similar to that of Soderblom (1970), Marcus (1970), and Hirabayashi et al. (2017). In this type of model, the 2-D spatial distribution of craters on the landscape is assumed to be well-represented by a 1-D size-frequency distribution (SFD). This 1-D SFD evolves in time due to the production of new craters and the degradation and removal of old craters. While the degradation of old craters is driven by changes in the 3-D topography, in a 1-D model it is assumed that the observability of a crater can be represented by a parameter that captures the average effect that degradation has on the visibility of craters in the population.

As we discussed in Section 1.3, the degradation of the small lunar maria craters in equilibrium can be modeled linear diffusive degradation. The basic principle of this idea is illustrated Figure 3. Therefore the basic component of our model is the linear diffusion equation given by equation (6), which relates topographic evolution to a parameter called the degradation state, $K$, which has units of m$^2$. The key difference between our analytical model and previous models, such as those by Soderblom (1970), Marcus (197), and Hirabayashi et al. (2017) is that we will model the degradation of craters and their ability to be recognized by a human crater counter in terms of a spatial average of their diffusive degradation state, $K$.

It is important to note that the actual morphological evolution of the landscape is not as simple. From equation (6), the change in the actual morphology of each crater, given by $h(x, y)$, as it is degraded diffusively depends on the topographic curvature, $\nabla^2 h$. This means that the sharp rims of craters degrade faster than the flat inner bowls, and that small craters degrade faster than large craters. However, because we are modeling linear diffusion, we can model degradation from multiple different craters and multiple processes within individual craters as a linear accumulation of $K$. By casting our model in terms of degradation state, we can eliminate much of the complexity involved in modeling the actual morphology of the surface.

We begin our development by defining two input functions, each of which is defined in terms of the degradation state $K$. We call these the *visibility function* and the *degradation function*. The visibility function, given by $K_v(r)$, quantifies the amount of accumulated diffusive degradation required to fully degrade a crater (i.e. degrade to the point that the crater is no longer recognizable) of radius, $r$. We demonstrate this concept in Figure 3, in which we have represented the range of degradation states of each crater at each time as a value between $0 < K < K_v$. In contrast, the *degradation function*, given by $K_c(\check{r} | \rho / \check{r}, \phi)$ quantifies the how much diffusive degradation a crater of a given size adds to the pre-existing landscape over some finite region of the surface. Here, $\check{r}$ is the radius of the crater that adds to the degradation state of the surface, and $(\rho, \phi)$ are polar coordinates with their origin at the crater center. The degradation function is an arbitrary scalar field function that could, in principle, capture the full range of spatial complexity shown in Figure 4A. Later, we will use a much simpler model in which the degradation region has a uniform value of $K_d(\check{r})$ over a region of radius $f_e\check{r}$, as shown in shown in Figure 4B.

In Section 2.1 we develop constraints on the visibility function, and capture some of the human factors of crater counting with a human crater counter calibration study. We next develop degradation functions that capture the processes involved in cratering in terms of the change over time of the diffusive degradation state (the degradation rate) of the surface as a function of cratering rate. In Section 2.2 we consider a model in which the degradation rate is constant. A constant diffusive degradation rate, $K'$, or equivalently a constant diffusivity, $\kappa$, which does not depend on the size scale of surface features is the underlying assumption of observational studies of the topographic evolution of lunar landscape, such as Craddock and Howard (2000) and Fassett and Thomson (2014). It is important to note that even if the degradation rate does not depend on scale, the evolution of landscape features does. The same value of the degradation state, $K$, will cause more morphological changes to small craters than large ones. We will show that a constant degradation rate model approximates the evolution of a surface in which the flux of primary micrometeoroids dominates diffusive degradation.

In Section 0 we show that the slope of the observed equilibrium SFD of of $\beta \sim 2$ requires a scale-dependence in the degradation rate. We show that a model in which each crater contributes to the degradation state in an amount proportional to its size naturally leads to the observed equilibrium slope for small simple lunar craters. Finally, in Section 3 we test our analytical models using the CTEM numerical code. Because equilibrium crater counts are most clearly seen on maria terrains for $r < 50$ m craters, we use the crater counts of the Apollo 15 landing site shown in Figure 2 as our primary constraint.

## 2.1. Defining and modeling the visibility function.

Consider a crater of radius $r$ that degrades diffusively according to equation (3). We will make the assumption that the spatial variability in the topographic diffusivity $\kappa$ across crater is relatively small, such that we can model the average diffusive degradation state of the crater in terms of a single value of $K$, using equation (6). This assumption was also made in Craddock and Howard (2000) as well as Fassett and Thomson (2014). We will later show numerically that this is a valid assumption for lunar mare craters. At some point $K$ can reach a value at which the crater is no longer recognizable by a human as a countable crater (see Figure 3). We can define a function $K_v(r)$, which describes the maximum degradation state that a crater can undergo before it becomes uncountable. We call this the *visibility function*.

In general, the amount of diffusive degradation a crater can accumulate before becoming uncountable will depend on size; the complete obliteration of a larger crater requires larger amounts of accumulated degradation than does a smaller one. To account for this scale dependence, we define the visibility function in terms of the degradation state at which the crater is no longer recognizable in the form of a power law function of crater radius as:

$$K_v(r) = K_{v,1} r^\gamma. \tag{7}$$

The visibility function given by (7) is defined in terms of the diffusive degradation state, and so has the units of length squared. This means that exponent $\gamma = 2$ is a special case in which the coefficient $K_{v,1}$ is dimensionless, and so represents the condition of geometric similarity. It arises when craters of different sizes have the same morphology, and so an image of a crater provides no information about its absolute scale. As an example, the complete erasure of a 100 m crater will require $100 \times$ as much accumulated degradation, $K$, as a 10 m crater, as long as the initial shape of

10 m craters is the same as that of 100 m craters.

Through experimentation we find that the visibility function for simple craters is constrained by two parameters: the initial depth-to-diameter ratio of the crater $(d/D)_{initial}$, and the minimum depth-to-diameter ratio of the craters that can be counted by a human, which we call $(d/D)_{cutoff}$. Simple craters on the lunar maria with $r > 200$ m typically begin with $(d/D)_{initial} \simeq 0.20 - 0.24$ (Fassett and Thomson, 2014; Pike, 1977; Stopar et al., 2017). This corresponds to an initial condition for the degradation state of the crater of $K = 0$.

To determine $(d/D)_{cutoff}$, and thus the degradation state where $K = K_v$ for a given crater of radius $r$, we performed a crater count calibration test using a human to count craters generated by CTEM. We used CTEM to generate a fictitious heavily-cratered terrain composed of simple craters. An image of the terrain was generated assuming a $45°$ solar incidence. Co-author Bryan Howl was first trained to count craters using the terrains counted in the study of Robbins et al. (2014). After training, Howl's crater count SFDs were close to the ensemble median of the Robbins et al. study, and hence, close to Fassett's counts. Howl was next tasked with counting craters on the CTEM-simulated surface. We then correlated the set of craters he identified, as well as the set that were produced in the simulation but not identified, with their depth-to-diameter. We found that the best fit value for $(d/D)_{cutoff} = 0.050$.

The visibility function is defined as the degradation state at the time that the craterform erodes such that its measured depth-to-diameter is $(d/D)_{cutoff}$. We next performed a simulation of the evolution of a simple crater subjected to diffusive degradation, given by equation (6). In this simulation, we generated a single crater in CTEM and degraded it with a simple constant linear diffusion model. We plot the resulting depth-to-diameter $(d/D)$ as a function of scale-normalized degradation state $(K/r^2)$ in Figure 5.

We used least squares fitting to fit the results shown in Figure 5 with an analytical function. We found that the value of $d/D$ as a function $K$ could be approximated by:

$$\frac{d}{D} = \frac{a^2}{(K/r^2 + b)^2}, \tag{8}$$

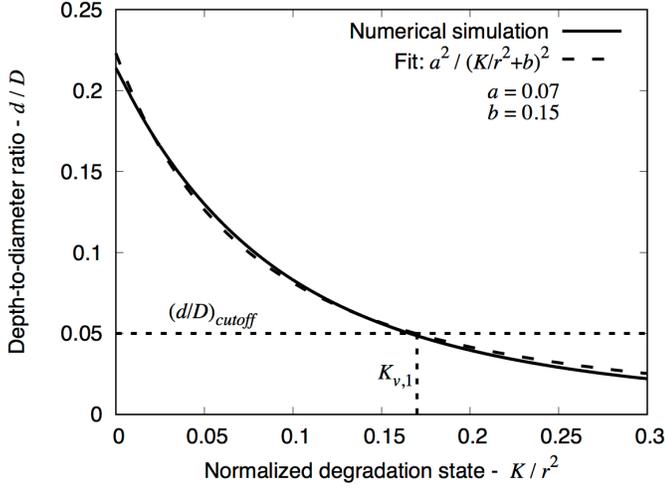

Figure 5. We show the results of a CTEM simulation of the depth-to-diameter ($d/D$) evolution of a simple crater undergoing linear diffusive degradation, given by equation (6). The accumulated degradation state, $K$, has been normalized by the $r^2$. The solid black line shows numerical results, while the long-dash line shows the fitted function given equation (8). The horizontal dash line shows the value of $(d/D)_{cutoff} = 0.050$ obtained from crater count calibration done by co-author Howl, and the vertical dash line shoes the corresponding value for the visibility function coefficient of $K_{v,1} = 0.17$.

where $a = 0.07$ and $b = 0.15$. The simulated $d/D$ as a function of $K/r^2$ as well as equation (8) are shown in Figure 5. This function is somewhat arbitrary, but it has the advantage of being easily inverted to obtain a simple expression for $K$ as a function of $d/D$, which we will next use to generate a visibility function from the results of our crater count calibration study. We invert equation (8) and write the degradation state associated with a particular depth-to-diameter ratio:

$$K = \left[a(d/D)^{-1/2} - b\right]r^2 \quad (9)$$

The coefficient $b$ in this fit is equivalent to $a(d/D)_{initial}^{-1/2}$, and therefore we can use equation (9) to write our visibility function $K_v(r)$.

$$K_v(r) = 0.07\left[(d/D)_{cutoff}^{-1/2} - (d/D)_{initial}^{-1/2}\right]r^2. \quad (10)$$

In CTEM, based on the way the code measures it, $(d/D)_{initial} = 0.218$ for simple craters. From our crater count calibration study $(d/D)_{cutoff} = 0.050$, and therefore from equation (10) then $K_{v,1} = 0.17$ and $\gamma = 2$. The strictest interpretation of this visibility function is that it quantifies the value of the degradation state of the most degraded, but still countable, simple craters on a well-resolved image of a lunar mare-like terrain that is illuminated at 45° solar incidence angle and is counted by co-author Bryan Howl.

There are a number of factors that can influence the visibility function for a given crater, including both morphological properties of the crater itself as well as the factors that contribute to crater recognition, such as the method used to identify the crater, the quality of the imagery, lighting conditions, and subjective human judgement. We will next examine the visibility function definition for insight into how these processes influence it.

If the visibility of a crater depends on its depth-to-diameter ratio, as in equation (10), then we should expect that $\gamma \neq 2$ if the initial depth-to-diameter ratio depends on crater radius, as is the case for complex craters (Kalynn et al., 2013; Pike, 1977) and possibly also on small simple craters (Daubar et al., 2014; Stopar et al., 2017). Many factors can influence the initial morphology of craters, and hence the visibility function. Craters that form on steeper pre-existing slopes should require less degradation to become uncountable than those formed on flat ground. Shallower crater should also degrade more readily than deeper craters. Evidence suggest that the depth-to-diameter may decreases to ~0.11–0.17 for simple craters with $r < 200$ m (Stopar et al., 2017). We can create a form of the visibility function to account for this effect. If we take $(d/D)_{initial} = 0.11$ for $r = 5$ m and $(d/D)_{initial} = 0.218$ for $r = 50$ m, then, applying equation (10) results in a visibility function with parameters $K_{v,1} = 0.073$ and $\gamma = 2.2$. Because equation (10) was derived from a study of simple crater counting, it likely does not apply to complex craters. As our study is focused on simple lunar mare craters, an investigation of the visibility function of complex craters is beyond the scope of this work.

The illumination of the image should also influence the visibility function. Heavily degraded craters are more easily seen at higher solar incidence angles where shadowing is more pronounced (Wilcox et al., 2005). For geometrically similar craters ($\gamma = 2$) the effect of this will be to change the coefficient $K_{v,1}$, where higher incidence angle (sun closer to the horizon) will result in a large value of $K_{v,1}$. The resolution of the image used to count craters should also influence the visibility function. In particular, it should add scale dependence such that the value of $K_v$ should drop as crater sizes approach the pixel scale of the image.

## 2.2. Crater equilibrium for the case of constant degradation rate

In this section we will derive an analytical model for the cumulative size-frequency distribution (SFD) of countable craters in equilibrium for the case where the rate of change of the degradation state of the surface is constant with respect to crater size, and is proportional to the rate of crater production.

By the definition of our dimensionless time unit, the rate of crater production is constant with respect to $X$. Therefore we express the degradation rate for this model as $K' = dK/dX$, where $K'$ is constant for all crater radii $r$ and has units diffusive degradation (m$^2$). Again, we make the assumption that any spatial variability in degradation state across the surface of interest is small, and therefore $K'$ does not depend on the position on the surface. We note that for the special case where the cratering rate is constant with time, such as lunar terrains less than $\sim 3$ Gy old, then $dX/dt$ is a constant and the degradation rate is proportional to the diffusivity as $K' = \frac{\kappa}{dX/dt}$.

Consider a surface that is populated at some rate with craters of radius $r$ over some differential range $dr$. As each new crater is added sequentially to the surface, all old craters are diffusive degraded at some degradation rate. As in Hirabayashi et al. (2017), we can define the accumulation and degradation of countable craters using a first-order linear differential equation:

$$\frac{d}{dX}\left(\frac{dn}{dr}\right) = \frac{d}{dX}\left(\frac{dn_p}{dr}\right) - k'\left(\frac{dn}{dr}\right), \qquad (11)$$

where $dn/dr$ is the differential number of countable craters, $dn_p/dr$ is the differential form of the production function, and $k'$ is a dimensionless degradation rate parameter that is defined as the fractional change in the differential number of countable craters per dimensionless time unit $X$.

At some point, the oldest crater in the sequence will accumulate enough degradation that it becomes too degraded to recognize, and is no longer counted. Even though this old crater is lost, a new crater is still added, and the net number of craters remains the same, e.g. $\frac{d}{dX}\left(\frac{dn}{dr}\right) = 0$. We illustrate this concept in Figure 3. Therefore the differential form of the equilibrium size-frequency distribution is given as:

$$\frac{dn_{eq}}{dr} = \frac{1}{k'}\frac{d}{dX}\left(\frac{dn_p}{dr}\right). \qquad (12)$$

The right-hand of equation (12) contains the differential form of the production size-frequency distribution. Cumulative size-frequency distributions are defined such that $\frac{dn}{dr} = -\frac{dn_{>r}}{dr}$, such that $n_{>r} = \int_{\infty}^{r}\frac{dn}{dr'}dr'$ (see Crater Analysis Techniques Working Group et al., 1979; Hirabayashi et al., 2017), and therefore using our definition of the production function from equation (1), we can write as:

$$\frac{d}{dX}\left(\frac{dn_p}{dr}\right) = n_{p,1}\eta r^{-\eta-1}. \qquad (13)$$

Substituting (13) into (12) gives us:

$$\frac{dn_{eq}}{dr} = \frac{n_{p,1}\eta}{k'}r^{-\eta-1}. \qquad (14)$$

Next we require an expression for the dimensionless degradation rate parameter, $k'(r)$. The visibility function is defined as the maximum degradation state a countable crater can have, and therefore we can define our degradation rate parameter $k'(r)$ in terms of the visibility function Section 2.1 as $K_v = K_{v,1}r^{\gamma}$. The visibility function depends on crater radius through the exponent $\gamma \sim 2$. In the model we are considering in this section, the absolute rate of degradation, $K'$, does not depend on crater radius. In other words, all craters are diffusively degraded at the same rate, but, as expressed by the visibility function, larger craters take longer to fully degrade than small craters.

We define the dimensionless degradation rate parameter in terms of diffusive degradation state as:

$$k'(r) = \frac{K'}{K_v(r)}. \qquad (15)$$

For the model under consideration here, $K'$ does not depend on crater radius, and therefore:

$$k'(r) = \frac{K'}{K_{v,1}} r^{-\gamma}. \tag{16}$$

We now combine equations (14) and (16) to write:

$$dn_{eq} = n_{p,1} \frac{K_{v,1}}{K'} \eta r^{-(\eta - \gamma) - 1}. \tag{17}$$

From our definition of the cumulative SFD, $n_{eq,>r} = -\int dn_{eq}$, and therefore

$$n_{eq,>r} = n_{p,1} \frac{K_{v,1}}{K'} \left( \frac{\eta}{\eta - \gamma} \right) r^{-(\eta - \gamma)}. \tag{18}$$

This model predicts that when the diffusivity does not depend on crater radius, such that diffusive degradation accumulates at the same rate as cratering, the exponent of the cumulative equilibrium SFD depends on the slope of the production function and the visibility function as $\beta = \eta - \gamma$.

For our Apollo 15 study area shown in Figure 2, the slope of the production cumulative SFD is $\eta = 3.2$. For geometrically similar simple craters the slope of the visibility function is $\gamma = 2$. Therefore, if crater degradation was dominated by a process that had a constant diffusivity, the cumulative equilibrium SFD would $\beta = 1.2$, rather than $\beta \sim 2$, as is observed (see Figure 2).

This suggests that the dominant degradation process that determines simple crater equilibrium cannot be modeled with a constant degradation rate that does not depend on scale. This means that, for the small lunar craters in equilibrium, the absolute degradation rate of a crater must depend on its size.

## 2.3. Crater equilibrium for the case of crater size-dependent degradation

In Section 2.2 we showed that a constant, crater size-independent degradation rate, $K'$, for terrains with steep sloped production populations results in an equilibrium SFD with a slope $\beta$ that depends on the production function slope $\eta$, as given by equation (18). For the steep-sloped production function SFD of small lunar craters, a constant degradation rate should result in an equilibrium SFD slope significantly shallower than the observed value of $\beta \sim 2$. A major problem with the constant degradation rate $K'$ is that it introduces dimension into the equilibrium SFD, but the observed value of $\beta = 2$ means that dimension does not appear in the equilibrium SFD. This lack of dimensionality in the equilibrium SFD requires a crater degradation mechanism that balances crater production the same way at all size scales.

We will now consider a different kind of model in which the degradation rate is itself determined by the crater production function. Here the formation of each crater contributes to the degradation state of the surface in a size-dependent way, such that small craters contribute small amounts of degradation over a small area and large craters contribute large amounts of degradation over a larger area. This is inherently the same degradation model that was developed by Soderblom (1970), Marcus (1970), and Hirabayashi et al. (2017), but we now develop it in terms of the diffusive degradation state, $K$.

In this model craters are both topographic features that are subject to degradation as well as the agents of topographic degradation. As we develop our model it is important to distinguish between the two distinct roles that craters play (features vs. agents of degradation). To do this we adopt the notation system used by Hirabayashi et al. (2017) in which we refer to countable craters that are being subjected to degradation have radius $r$, while newly formed craters that contribute to degradation have radius $\check{r}$. Any individual crater only ever takes on the role of degradation agent ($\check{r}$) once (the moment it forms), and thereafter it becomes a topographic feature to be degraded ($r$). Therefore any individual crater will play both roles in the evolution of the terrain, and we will assume that the contribution to degradation by new craters will depend on their radius $\check{r}$.

Because the impact crater production function for the mare terrains in this study is well-fit by a power law cumulative SFD, the larger craters occur with lower probability than smaller ones. Therefore, for a piece of the surface of a given area, the longer it accumulates craters, on average the larger will be the size of the largest crater that contributes to its degradation. Similarly, for a given interval of time (or total crater accumulation) the larger the area under consideration, on average the larger will be the size of the largest crater that contributes to the degradation of that area. This time and spatial-scale dependence on the degradation rate results in anomalous diffusion, rather than classical diffusion (Li and Mustard, 2000; Vlahos et al., 2008), and is a key constraint, along with geometric similarity, on the conditions that lead to the equilibrium SFD slope value of $\beta = 2$.

Soderblom (1970) developed a model for topographic diffusivity that accounted for the size-scale dependence of diffusive degradation. In Soderblom's model, slope-dependent distribution of the proximal ejecta blankets of small craters was thought to be the primary degradation mechanism for large craters. That is, when a small crater forms inside the wall of a large crater, its proximal ejecta blanket will contain more mass on the downslope side than the upslope side. When averaged over many impact events, this slope-dependent asymmetry in the mass distribution of small crater ejecta blankets will naturally result in topographic diffusion of the larger crater as expressed in equation (4).

We may rewrite the diffusivity expression given in equation (14) of Soderblom (1970) using our notation system as:

$$\kappa = C_1 \frac{n_{p,1>r} \check{r}_{max}^{4-\eta}}{4-\eta}, \tag{19}$$

where $C_1 \sim \frac{1}{2}$ is a constant and $n_{p,1>r}$ is cumulative number of produced craters greater than 1 m in radius. The term $\check{r}_{max}$ accounts for the size-scale dependence of the diffusive degradation process resulting from cratering. In his model, $\check{r}_{max}$ is the radius of the largest new crater whose effects can be averaged over the old crater of radius $r$ over its lifetime. The constant $C_1$ is related to the mass distribution in the ejecta thickness, with $C_1 \approx \frac{1}{2}$ for ejecta thickness profiles roughly similar to what is observed on the lunar surface (McGetchin et al., 1973). A challenge faced by Soderblom (1970) was to determine the appropriate value of $\check{r}_{max}$ for the problem of equilibrium of small maria craters. He was able to fit their modeled equilibrium SFD to the observed empirical equilibrium SFD for $r < 50$ m lunar maria craters, but his results were very sensitive to a number of assumptions that were difficult to constrain from observations.

Here we will develop a model similar for diffusive degradation of larger craters from the formation of smaller craters similar to that of Soderblom (1970), as well as that of Ross (1968). We consider a generic model in which each new crater of size $\check{r}$ causes some amount of diffusive degradation to the pre-existing landscape over some finite region scaled by the crater size. We will model this per-crater contribution to the degradation state of the surface using a *degradation function*.

An impact is a complicated event involving, among other things, the creation of a bowl-shaped depression and raised rim, transport of ejecta, and the generation of seismic waves (see Figure 4A). The degradation function quantifies how all the processes involve in cratering contribute to the diffusive degradation state of the surface over an extended area. Because the degradation state is defined in terms of topographic diffusion, the changes to the topography depend on the local slope via equation (6).

Each new crater of radius $\check{r}$ contributes a finite amount to the degradation state of the surface, $K$. This contribution is not spatially uniform across the surface, but occurs over some region $R_d$ with surface area $A_d$. Consider a degradation function in the form of a scalar field function $K_c(\check{r} | \rho/\check{r}, \phi)$, where $\check{r}$ is the radius of the new crater that is contributing to degradation, and $(\rho, \phi)$ are the polar coordinates of a point on the surface with respect to the center of the new crater.

We can define a function $K_d$ that only depends on crater radius $\check{r}$ and a non-dimensional scale factor $f(\rho/\check{r}, \phi)$ such that:

$$K_c(\check{r} | \rho/\check{r}, \phi) = K_d(\check{r}) f(\rho/\check{r}, \phi). \tag{20}$$

The contribution to the degradation state of the surface by each new crater of radius $\check{r}$ is:

$$K(\check{r} | R_d) = \frac{K_d(\check{r})}{A_d/\check{r}^2} \iint_{R_d} f(\rho/r_j, \phi) \, \rho \, d\rho \, d\phi. \tag{21}$$

The field function $f(\rho/\check{r}, \phi)$ is known as an intensity function, and it can be arbitrarily complex, as it can represent any process involved in the formation of the new crater that gives rise to, or can be approximated by, linear topographic diffusion in the form of equation (6). For instance, it can represent the slope-dependent proximal ejecta mass asymmetry of both the primary crater, as in Soderblom (1970), as well as that due to secondary craters. It can also represent seismic shaking, which was modeled as a topographic diffusion process in Richardson (2009). The diffusive intensity function can also potentially have complex spatial heterogeneity if the high energy deposition of distal ejecta, such as secondary craters or ballistic sedimentation, are organized into ray-like features (Elliott et al., 2018; Huang et al., 2017; Oberbeck, 1975).

In this section we are developing an analytical model that is merely a representation of the visible crater SFD, and which contains no spatial information. For the purposes of our analytical model, we can average out all of the spatial complexities of the degradation function by considering an equivalent problem in which the diffusive degradation due to each crater is uniform over a circle of radius $f_e\check{r}$. The scale factor $f_e$ is found using a surface integral of the non-dimensional scalar field function $f$ over all possible spatial points:

$$f_e = \sqrt{\int_0^{2\pi} \int_0^\infty f(\xi,\phi)\xi \, d\xi \, d\phi}. \quad (22)$$

This defines a circle of radius $f_e\check{r}$ for which the degradation contribution is uniform and has the same spatially averaged degradation contribution $K$ as the original, more complex field function did (see Figure 4B). This approach averages out all of spatially-dependent complexities in the ways in which the impact process degrades the landscape, and so while it simplifies the math considerably, it may miss some process that are important in the development of the surface. We will explore the importance of capturing this spatial heterogeneity in Section 3 when we model the equilibrium process numerically with the CTEM Monte Carlo code.

Due to the stochastic nature of cratering, there is a finite probability that each old countable crater of radius $r$ will experience a single degradation event with $\delta K > K_v(r)$. Cookie-cutting is an example of such an event. The fact that the empirical crater equilibrium cumulative SFD is relatively similar across multiple locations on the lunar surface (e.g. Xiao and Werner, 2015) implies that the degradation of any particular crater is unlikely to be large relative to its visibility function, and that the degradation state $K$ at any point on the surface is approximately equal to the ensemble-mean $K$ and spatially-average $K$, and that these are equal.

Recall that we have previously defined the production and equilibrium crater cumulative SFDs in equation (1) and (2) as power laws with slopes for the equilibrium and production functions of $\beta$ and $\eta$, respectively. We also defined the visibility function in Section 2.1, equation (7), which we model as a power law with slope $\gamma \sim 2$. Because our constraint (the empirical equilibrium SFD) is well characterized by a power law, and all other inputs are also power laws, it is reasonable to assume that our per-crater degradation function should also be a power law. We define our per-crater degradation function as:

$$K_d(\check{r}) = K_{d,1}\check{r}^\psi. \quad (23)$$

Just as we discussed in Section 2.1 in developing the visibility function, the degradation function is defined in terms of the degradation state, $K$, which has units of m², and therefore $\psi = 2$ represents a special case in which $K_{d,1}$ contains no information about absolute scale. As long as $f_e$ also does not depend on $\check{r}$, then a degradation function slope of $\psi = 2$ is a one that exhibits geometric similarity.

The amount of degradation contributed by each new crater is given by the degradation function $K_c(\check{r}|\rho/\check{r},\phi)$ applied over some region surrounding the newly-formed crater. We can use the equivalent uniform circular degradation region concept shown in Figure 4B, and take the degradation function as uniform $K_d(\check{r})$ over a circular region with area $A_d = \pi f_e^2 \check{r}^2$.

As in Section 2.2, we begin with our equation for equilibrium given by equation (14) as $\frac{dn_{eq}}{dr} = \left(\frac{n_{p,1}\eta}{k'}\right)r^{-\eta-1}$, where $k'$ is again the dimensionless degradation rate parameter. Unlike in the case we investigated in Section 2.2 where the degradation rate parameter $k'$ was a function of the constant diffusivity and the visibility function, here the degradation rate parameter arises from the collective accumulation of degradation from all new craters of size $\check{r}$ over time. We can therefore define the dimensionless degradation rate parameter as $k' = \int dk'$, where $dk'$ is the differential contribution to the complete degradation of old crater $r$ by new crater $\check{r}$ per unit time.

When the degradation from a new crater is small relative to the visibility function of the old crater, the new crater contributes partial degradation to the old craters. Because we are averaging over all craters on the surface, these partial degradations are treated as a fractional loss of total countable crater number.

However, because a crater can only ever be lost once, we must take care not to over-count the loss of craters in this averaging. This transition occurs when the degradation function of the new crater is equal to the visibility function of the old crater, or $K_d(\check{r}) = K_v(r)$. Therefore we identify two regimes of degradation that depend on the relative magnitude of the per-crater degradation function and the visibility function of the crater being degraded. The boundary between these two regimes occurs at radius $r_{crit}$ given as:

$$r_{crit} = \left(\frac{K_{v,1}}{K_{d,1}}\right)^{1/\psi} \check{r}^{\gamma/\psi}. \tag{24}$$

While many of our underlying assumptions are different, our two regimes are mathematically analogous to the two regimes identified by Hirabayashi et al. (2017). In our model, $\check{r} < r_{crit}$ is equivalent to what Hirabayashi et al. called the "sandblasting" regime, and $\check{r} > r_{crit}$ is equivalent to what they called the "cookie-cutting" regime (see their Figure 5). However our $\check{r} > r_{crit}$ regime is not the same as true cookie-cutting, as it does not require a crater is lost by direct overlap. For instance, a small crater completely buried in proximal ejecta would fall in this regime. We therefore call these two regimes "partial degradation" ( $\check{r} < r_{crit}$ ) and "total obliteration" ($\check{r} > r_{crit}$).

Consider an old crater of radius $r$ that is within the degradation region of a new crater of radius $\check{r}$. If this pair of craters is in the partial degradation regime, $\check{r} < r_{crit}$, the old crater is partially degraded by the new crater. The amount of partial degradation is the ratio of the spatially averaged degradation contribution of the new crater to the visibility function of the old crater, which is expressed as:

$$dk'_{\check{r}<r_{crit}} = \frac{K_d(\check{r})}{K_v(r)} A_d\, dn_p(\check{r})$$
$$= -\frac{K_{d,1}}{K_{v,1}r^\gamma}\pi f_e^2 n_{p,1}\eta \check{r}^{\psi-\eta+1} d\check{r}, \tag{25}$$

If instead this pair of craters is in the total obliteration regime, $\check{r} > r_{crit}$ , the old crater is completely obliterated, which is expressed in terms of the spatially averaged degradation contribution as:

$$dk'_{\check{r}>r_{crit}} = \pi f_e^2 n_{p,1}\eta \check{r}^{-\eta+1} d\check{r}. \tag{26}$$

There is likely to be a smooth transition between the two regimes near $r_{crit}$, but for simplicity we will assume the width of this transition region is negligible.

Using our regime definitions, we write our complete dimensionless degradation parameter as:

$$k' = \frac{1}{K_{v,1}r^\gamma}\int_0^{r_{crit}} K_{d,1}\pi f_e^2 n_{p,1}\eta \check{r}^{\psi-\eta+1} d\check{r}$$
$$+ \int_{r_{crit}}^{\infty} \pi f_e^2 n_{p,1}\eta \check{r}^{-\eta+1} d\check{r}. \tag{27}$$

The lower integral defines the regime of partial degradation, and the upper integral defines the regime of total degradation. The solution to this is

$$k' = n_{p,1}\eta\pi f_e^2 \left(\frac{K_{v,1}}{K_{d,1}}\right)^{-(\eta-2)/\psi}$$
$$\times \left[\frac{\psi}{(\eta-2)(\psi-\eta+2)}\right] r^{-\frac{\gamma}{\psi}(\eta-2)}, \tag{28}$$

as long as $2 < \eta < \psi + 2$, or otherwise one of the two limits would yield an infinite result. For shallow production SFDs (i.e. $\eta < 2$), this is the "cookie-cutting dominated" regime. Hirabayashi et al. (2017) showed that the equilibrium SFD in this regime has the slope $\beta = \eta$. For very steep production SFDs (i.e. $\eta > \psi + 2$), the smallest craters dominate the degradation, and such steep SFDs must "roll over" to a shallower slope at some small size, otherwise craters of any size would degrade at a faster rate than their production rate. This result was also found by Soderblom (1970) when $\eta > 4$, with a model for which an implied per-crater degradation function was geometrically similar ($\psi$=2).

Substituting (28) into (14) we have:

$$\frac{dn_{eq}}{dr} = \left(\frac{1}{\pi f_e^2}\right)\left(\frac{K_{v,1}}{K_{d,1}}\right)^{\frac{(\eta-2)}{\psi}}$$
$$\times \left[\frac{(\eta-2)(\psi-\eta+2)}{\psi}\right] r^{-\eta+\frac{\gamma}{\psi}(\eta-2)-1} \tag{29}$$

After some simplification and integration, this yields a solution for the cumulative equilibrium SFD of:

$$n_{eq,>r} = \left(\frac{1}{\pi f_e^2}\right)\left(\frac{K_{v,1}}{K_{d,1}}\right)^{\frac{(\eta-2)}{\psi}}$$
$$\times \left[\frac{1-(\eta-2)/\psi}{\eta/(\eta-2)-\gamma/\psi}\right] r^{-\left[2\frac{\gamma}{\psi}-\eta\left(\frac{\gamma}{\psi}-1\right)\right]}, \tag{30}$$

which has the form $n_{eq,>r} = n_{eq,1}r^{-\beta}$ with a slope given as:

$$\beta = 2\frac{\gamma}{\psi} - \eta\left(\frac{\gamma}{\psi} - 1\right). \qquad (31)$$

The parameters that set the equilibrium slope are the slope of the visibility function ($\gamma$), the slope of the per-crater degradation function ($\psi$), and the slope of the production function ($\eta$). For geometrically similar craters with geometrically similar degradation contribution we have $\psi = 2$ and $\gamma = 2$. Therefore in geometrically similar case, the dependence on the production function slope $\eta$ vanishes and $\beta = 2$. This occurs as a consequence of the fact that the production function appears in both the accumulation and degradation terms of equation (11). In other words, for surface with an equilibrium SFD slope $\beta = 2$, the production and destruction of craters must be intrinsically linked. The population of counted craters must originate in a production function with the same slope and cratering rate as that of the craters controlling their degradation. This also implies that if degradation were dominated by a process that is independent of the crater production then $\beta = 2$ would be unlikely.

This in contrast with the predictions for equilibrium using a constant degradation rate model derived in Section 2.2 where $\beta = \eta - \gamma$. In that case, for $\eta = 3.2$, $\beta = 1.2$ for geometrically similar craters. For the crater size-dependent degradation model, when the visibility and per-crater degradation function are not strictly geometrically similar, but are still close ($\gamma \approx \psi$), then the equilibrium slope $\beta$ only weakly depends on the production function slope, $\eta$, rather than being linearly dependent as in the constant degradation rate model. The crater size-dependent degradation rate model is therefore more consistent with observations of small simple crater equilibrium of the lunar maria than the constant degradation rate model.

Table 1. Definitions of terms used in this text, including the section in the text where the terms are introduced and defined.

| Parameter | Name | Units | Section |
|---|---|---|---|
| $r$ | Old crater radius | m | 1.1 |
| $\check{r}$ | New crater radius | m | 0 |
| $n_{p,>r} = n_{p,1}Xr^{-\eta}$ | Production function | #/m$^2$ | 1.1 |
| $n_{p,1}$ | Production function coefficient | m$^{\eta-2}$ | 1.1 |
| $\eta$ | Production function slope | - | 1.1 |
| $X$ | Dimensionless time | - | 1.1 |
| $n_{eq,>r} = n_{eq,1}r^{-\beta}$ | Equilibrium SFD | #/m$^2$ | 1.1 |
| $n_{eq,1}$ | Equilibrium coefficient | m$^{2-\beta}$ | 1.1 |
| $\beta$ | Equilibrium slope | - | 1.1 |
| $n_{gsat,>r} = n_{gsat,1}r^{-2}$ | Geometric saturation SFD | #/m$^2$ | 1.1 |
| $n_{gsat,1} = 0.385$ | Geometric saturation coefficient | - | 1.1 |
| $\kappa$ | Topographic diffusivity | m$^2$/y | 1.3 |
| $K$ | Topographic degradation state | m$^2$ | 1.3 |
| $K' = dK/dX$ | Degradation rate | m$^2$ | 2.2 |
| $k'$ | Dimensionless degradation rate | - | 2.2 |
| $K_v = K_{v,1}r^{\gamma}$ | Visibility function | m$^2$ | 2.1 |
| $K_{v,1}$ | Visibility function coefficient | m$^{2-\gamma}$ | 2.1 |
| $\gamma$ | Visibility function slope | - | 2.1 |
| $K_d = K_{d,1}\check{r}^{\psi}$ | Per-crater degradation function | m$^2$ | 0 |
| $K_{d,1}$ | Degradation function coefficient | m$^{2-\psi}$ | 0 |
| $\psi$ | Degradation function slope | - | 0 |
| $A_d = \pi f_e^2\check{r}^2$ | Degradation region area | m$^2$ | 0 |
| $f_e$ | Degradation size scale factor | - | 0 |

Finally, we can write equation (30) in terms of the uniform circular degradation function parameters:

$$
K_{d,1} r^{\psi}
= K_{\nu,1} \left\{ \pi f_e^2 n_{eq,1} \left[ \frac{\gamma \beta}{(\eta-2)(\beta+\gamma-\eta)} \right] \right\}^{-\frac{\gamma}{(\eta-\beta)}}
\times r^{\gamma \frac{(\eta-2)}{(\eta-\beta)}} \quad (32)
$$

We will use equation (32) to constrain the degradation function required to match the observed equilibrium SFD in Sections 3.3 and 3.4.

# 3. Testing the analytical models for crater equilibrium with CTEM

In Section 2 we developed 1-D diffusion-based analytical models for crater equilibrium. The functions that define the inputs and constraints used in our models, along with their units, are shown . Here we will test the analytical models using the Cratered Terrain Evolution Model (CTEM) and use constraints from both observations and numerical experiments to constrain the parameters that determine the equilibrium SFD. CTEM is a 3-D Monte Carlo landscape evolution code that models the topographic evolution of a surface that is subjected to impact cratering. It has previously been used to study the lunar highlands cratering record (Minton et al., 2015; Richardson, 2009) and impact transport of compositionally distinct surface materials (Huang et al., 2017). Because it is a three-dimensional code, it can readily model topographic diffusion. The simulations we perform with CTEM contain many of the complexities of the landscape that were assumed to average out in the 1-D analytical models developed in Section 2. Therefore the CTEM simulations can be used to test how robust were the assumptions that went into the development of the analytical models.

An important part of capturing the way that small craters degrade larger craters is to ensure that the ejecta generated by each crater has the correct slope-dependent mass distribution, as discussed by Ross (1968) and Soderblom (1970). In principle, this is relatively straightforward, as CTEM models the distribution of ejecta deposits from each crater using the ballistic trajectory of parcels of material emerging from the transient crater region. CTEM calculates the velocity and ejection angle of each parcel of ejecta and emplaces the ejecta downrange of the impact site using ballistic range equations assuming a flat plane geometry, as described in Richardson (2009). However, the ejecta thickness model in Richardson (2009) was based on a model for excavation flow, which scales in a complex way with crater size, and therefore does not exhibit geometric similarity. To simplify the comparisons between our numerical results and our analytical model, we needed to better control the relationship between the ejecta distribution and the size of the crater. In other words, we needed to ensure that our ejecta model exhibited geometric similarity.

First, we modified the model for the shape of simple craters from the original parabolic shape used in Richardson (2009). We now create simple craters that conform to the initial shape function given by equation (4) in Fassett and Thomson (2014). Next, we modified the ejecta model in CTEM so that the ejecta profiles for a crater that forms on a perfectly flat surface will follow the equation $h = h_{rim}(d/\check{r})^{-3}$, where the thickness $h$ of the deposit is a function of radial distance $d$ and crater radius $\check{r}$ (McGetchin et al., 1973). In our modified version of CTEM's ejecta blanket thickness model $h_{rim}$ is determined such that mass is conserved by the formation of the crater. Because of these modifications, craters and their ejecta blankets are geometrically similar at all sizes for this study, with one exception: for craters that form on slopes, we distort the ejecta distribution to preferentially deposit ejecta mass downslope using ballistic flight equations. The steeper the initial slope, the more of the ejecta is deposited on the downslope side, which gives rise to the diffusive degradation discussed in Soderblom (1970) (see Figure 6).

Ejecta blankets can contribute to the removal of craters through direct burial. Ejecta burial is not strictly a diffusive degradation process, however it has similar properties. The more buried a crater becomes, the harder it is to observe. As Fassett et al. (2011) showed, an old crater that is buried in ejecta from a new crater that is as thick as the depth of the old crater will not be countable. We will therefore approximate the modification of a crater by burial as a diffusive degradation process that depends on the local thickness of a new ejecta deposit. We will use the visibility function derived in Section 2.1 to quantify the amount of diffusive degradation that will remove craters in the same way as ejecta burial.

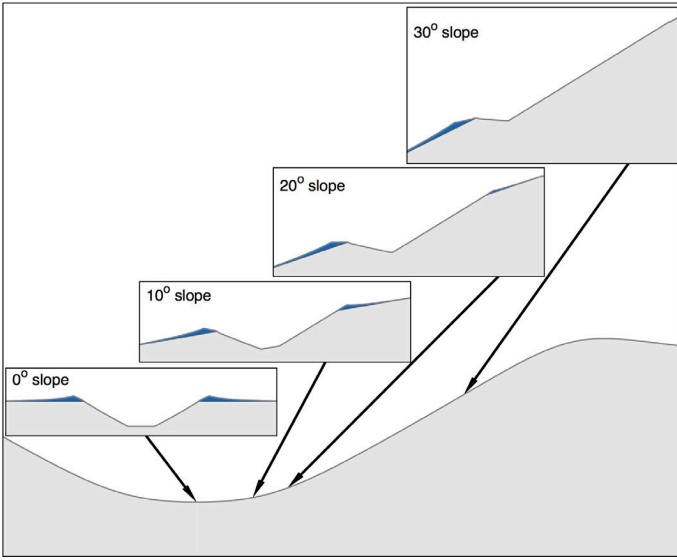

Figure 6. An illustration of the slope-dependent asymmetry in the proximal ejecta of small impacts into pre-existing topography. The large shaded region is a portion of the profile of a partially degraded large crater. Each inset shows the CTEM-generated profile of a crater that forms at the local slope indicated by the arrow. The dark blue regions in each inset shows the portion of the profile of each small crater that is the ejecta and raised rim. As the slope angle of the target surface increases, proportionally more of the ejecta is deposited downslope.

Minton et al. (2015) developed an ejecta blanket degradation model based on diffusion that erases craters whose depths are equal to the ejecta blanket thickness. The value of the diffusion parameter in that model was set through trial and error, but now we can formally define it using our degradation model. In this model, a crater is erased when the ejecta thickness is at least the thickness required to degrade a crater. In other words, $h_{ej} = 2r_{degraded}[(d/D)_{initial} - (d/D)_{cutoff}] = 0.34r_{degraded}$, where $r_{degraded}$ is the largest crater that is fully buried by an ejecta blanket with thickness $h_{ej}$. This means that

$$K_{d,ej} = K_v \left(\frac{h_{ej}}{r_{degraded}}\right)^2 = 1.5h_{ej}^2, \quad (33)$$

Our model for ejecta burial assumes that burial follows the diffusive degradation relationship of equation (6), rather than being linear with burial depth, which is not strictly correct. However as we show later, ejecta burial likely contributes very little to the degradation of small lunar craters in equilibrium, so the approximations we use do not become important for our major results.

Due to the finite resolution of CTEM, we must consider the effects of subpixel cratering. In our analytical models, we assumed that the production function SFD was continuous down to infinitesimal sizes. However, the NPF is only defined for $r > 5$ m craters. If we truncated the production function to craters this small, we may introduce non-physical behavior in the crater SFD from the lack of small impactors. We make a simple, but reasonable, assumption that the SFD continues with the same slope into the size range of micrometeoroid impactors. We extrapolate the SFD of the production function down to craters as small as $\sim 6$ μm. To handle the population of craters smaller than the resolution limit of 1 m in our simulation, we periodically apply diffusive degradation to each pixel. This subpixel diffusion is modeled using the same degradation function we use for the resolved craters, but instead of modeling each crater individually, we use the subpixel portion of the crater production function scaled to the pixel area. The differential form of the subpixel crater production function is used as an input into a Poisson random number generator, and then the resulting differential number of subpixel craters is multiplied by the degradation contribution using the degradation function. This is done on each pixel of the simulation domain.

In Section 3.1 we first perform a series of numerical experiments to constrain a degradation function using a model in which the dominant mechanism for diffusive degradation is the preferential downslope deposition of proximal ejecta by primary impactors. We will show that this model is inadequate to match the observed equilibrium SFD for simple lunar craters, and some additional source of diffusive degradation is required that was not accounted for in that model. In Section 3.2 we develop a production function that contains an enhanced micrometeoroid population, which approximates the constant degradation rate model that was developed in Section 2.2, and show that, as predicted by the analytical model, this type of model also does not match the observed equilibrium SFD. Next, we use the observed equilibrium SFD in order to constrain the properties of the required additional diffusive degradation for a uniform degradation region model in Section 3.3. In Section 3.4. we develop a model in which the dominant source of diffusive degradation occurs in a spatially heterogenous distal ejecta, which is crater size-dependent

### 3.1. CTEM simulation of the constant slope primary production function.

We use the capabilities of CTEM to test whether the observed equilibrium SFD is achieved for a steep-sloped production SFD when the only source of crater degradation is the primary production population, and the primary production population of small craters (micrometeoroids) is a simple power law extrapolation of the production function of the larger craters. We model the degradation by primary impactors using a slope-dependent proximal ejecta distribution model similar to that of Soderblom (1970). We only consider primary impactors and their proximal ejecta blankets. We also used the analytical model developed in Section 0 in order to predict the equilibrium cumulative SFD for this model. To do this, we first constrained the per-crater degradation function, $K_d(\check{r})$, for the proximal ejecta redistribution model.

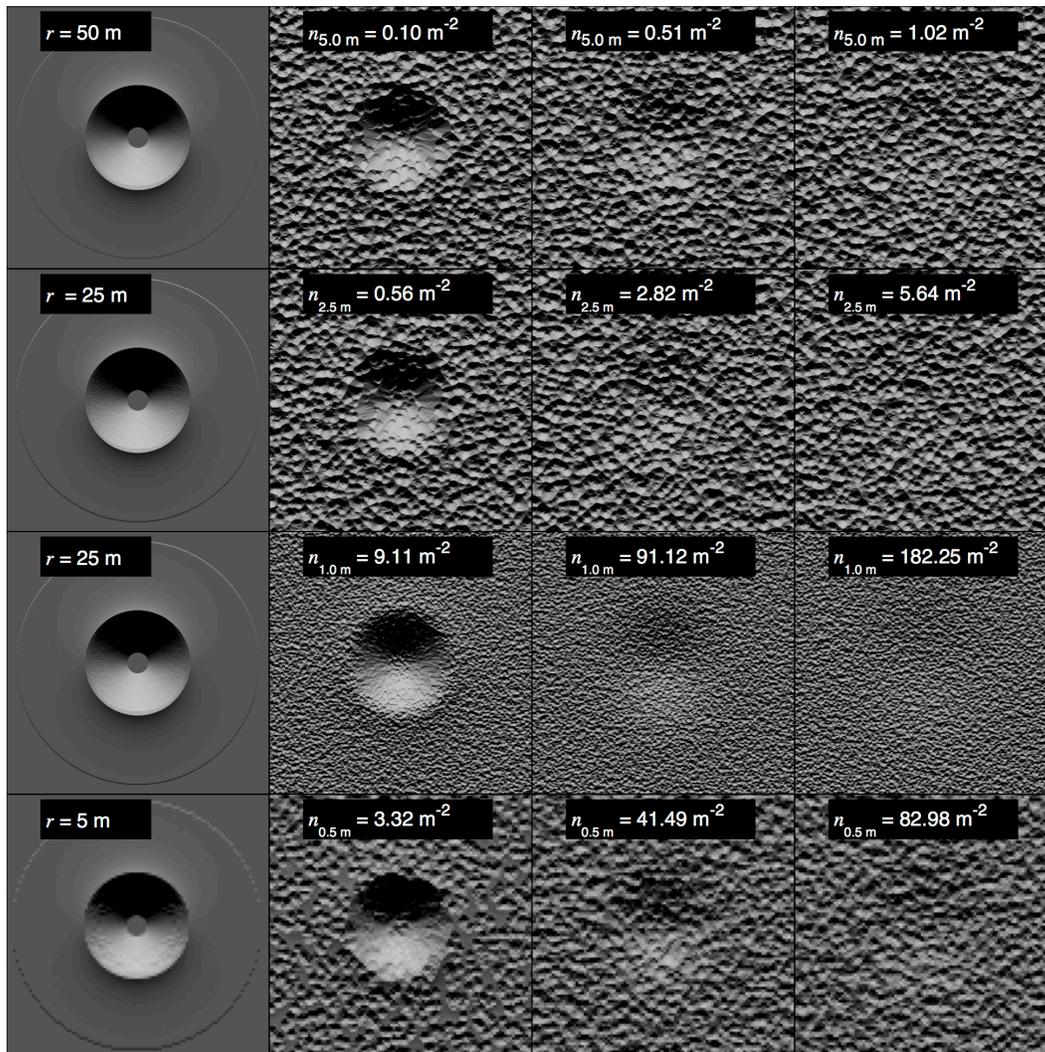

Figure 7. Shaded DEM maps from single crater population bombardment experiments that were used to determine the degradation function for the slope-dependent proximal ejecta mass distribution of primary impactors. All simulations have a resolution of 0.25 m/px. The large crater's ejecta blanket was truncated at 2.3$r$ (the edge of the continuous ejecta blanket) in order to prevent overlapping ejecta due CTEM's repeating boundary condition. The ejecta of small craters were truncated at 5.0$\check{r}$. No additional diffusive degradation was included in these simulations.

We use a production function with a constant slope of $\eta = 3.2$ for $r > 6$ μm craters.

We performed a set of simple numerical experiments to determine the per-crater degradation function $K_d(\check{r})$ that results from the slope-dependent downslope ejecta deposition model we included in CTEM. In these experiments we eroded a single large "test crater" by a population of smaller craters of identical size. Examples of the outputs of these simulations are shown in Figure 7 for different combinations of pixel resolution, test crater size, and total cratering. We computed the average profile of the large test crater from our DEMs at different points in the simulation.

These profiles were then fitted to the profile of the same size crater undergoing classical diffusion by equation (6) and solving for the required degradation state $K$ to match the simulated profiles. The values of $K_{d,1}$ and $f_e$ are degenerate when deriving the degradation function in this way, and so we assume $f_e = 1$. The results and a fit to the per-crater degradation function are shown in Figure 8. The results of these experiments also show the validity of the assumption that spatial variability in the mean degradation parameter is small.

Over the range of craters we considered in these experiments, our simulation results are well fit to a per-crater degradation function given by $K_d = 0.27\check{r}^2$. The coefficient, $K_{d,1} = 0.27$ tells us how the effectiveness, or power, of the diffusive degradation, and the exponent $\psi = 2$ is a consequence of the geometric similarity of crater morphology at all sizes in CTEM. Although we simulated the degradation due to craters between $\check{r} = 0.25 - 5$ m, our best fit exponent $\psi = 2$ indicates that the CTEM craters are geometrically similar at all scales, and the coefficient $K_{d,1}$ is unitless. The absolute scale in these simulations is therefore arbitrary, and our degradation function is applicable to any size crater, so long as we maintain similar crater geometry for all craters in our numerical simulations.

Using our degradation function derived from the CTEM numerical experiments, we can estimate the equilibrium crater SFD that we should expect using the analytical model developed in Section 0. We will compare our results with that for the Apollo 15 landing site, shown in Figure 2. Our production function parameters are $n_{p,1} = 4.3$ m$^{1.2}$ and $\eta = 3.2$. The visibility function parameters derived from our human crater count experiment are $K_{v,1} = 0.17$ and $\gamma = 2$. From our single-size crater experiments our degradation model has parameters $K_{d,1} = 0.27$, $\psi = 2$, and $f_e = 1$. Using equation (30) we estimate that $n_{eq,1} = 0.058$ and $\beta = 2$.

Considering only the primary production population with a constant SFD slope results in a value of $n_{eq,1}$ that is nearly an order of magnitude higher than Fit 1 value of 0.0084 from Figure 2. This suggest that if preferential downslope ejecta deposition of primary craters was the dominant mechanism for diffusive degradation on the lunar surface, then crater densities would be much higher than what is observed on terrains such as the Apollo 15 landing site. We tested this prediction using CTEM to simulate the Apollo 15 landing site using with a resolution of 1 m/pixel on a domain $1000 \times 1000$ pixels. We bombarded our simulated surface with a crater production function with $n_{p,1} = 4.3$ m$^{1.2}$ and $\eta =$

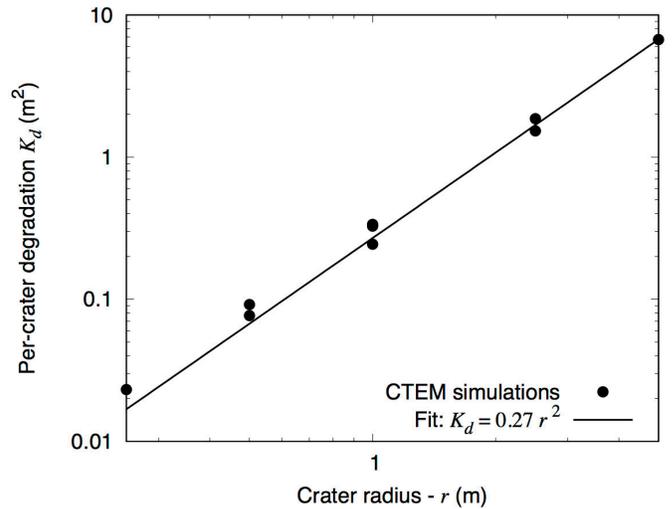

Figure 8. The degradation function model obtained from our single crater simulations, shown in Figure 7. The primary degradation process in these simulations is slope-dependent mass redistribution in the proximal ejecta of the small craters (see Figure 6). Ejecta burial is also modeled by applying equation (33) to the pre-existing terrain beneath each small crater's ejecta. Each point represents a fit to the per-crater contribution to the degradation state from each simulation. Multiple points for a given crater radius indicate simulations done at different pixel resolutions. The solid line is the best fit to the data and gives the parameters of the degradation function for this model of $K_{d,1} = 0.27$ and $\psi = 2$, assuming $f_e = 1$.

3.2. The results are shown in Figure 9.

Figure 9 shows the shaded DEM and crater counts at the end of our simulation of the slope-dependent ejecta deposition model. The simulation DEM is qualitatively much rougher in texture than the real Apollo 15 landing site at similar scales shown in Figure 2. In Figure 9 we plot both the crater counts of the real surface from co-author Fassett in Robbins et al. (2014) and the crater counts of our CTEM simulation. We also plot both the predicted equilibrium line from equation (30) and the geometric saturation line for reference. As we can see, the crater number density we predict from our analytical model as well as the crater number density we reach in the numerical simulation are far higher than is observed on the lunar surface. Both our analytical and numerical results suggest that our slope-dependent ejecta deposition model is inadequate for explaining the observed degree of diffusive degradation on the lunar surface.

In the next sections we turn the problem around and treat the empirical equilibrium SFD as a constraint and solve for the uniform circular degradation function parameters $K_{d,1}$ and $f_e$ that match that constraint. Once we determine what parameters best match observational constraints, we will then discuss the implications of our derived degradation function on what processes dominate diffusive degradation of the lunar surface.

### 3.2. CTEM simulation of a production function with enhanced micrometeoroids

In Section 3.1 we used a numerical simulation in CTEM to show that if diffusive degradation is dominated by slope-dependent proximal ejecta distribution, the resulting equilibrium cumulative SFD was an order of magnitude higher than is observed. Our numerical results were consistent with what we predicted using the analytical model developed in Section 0. In order to match the observed equilibrium SFD, we require some additional source of diffusive degradation.

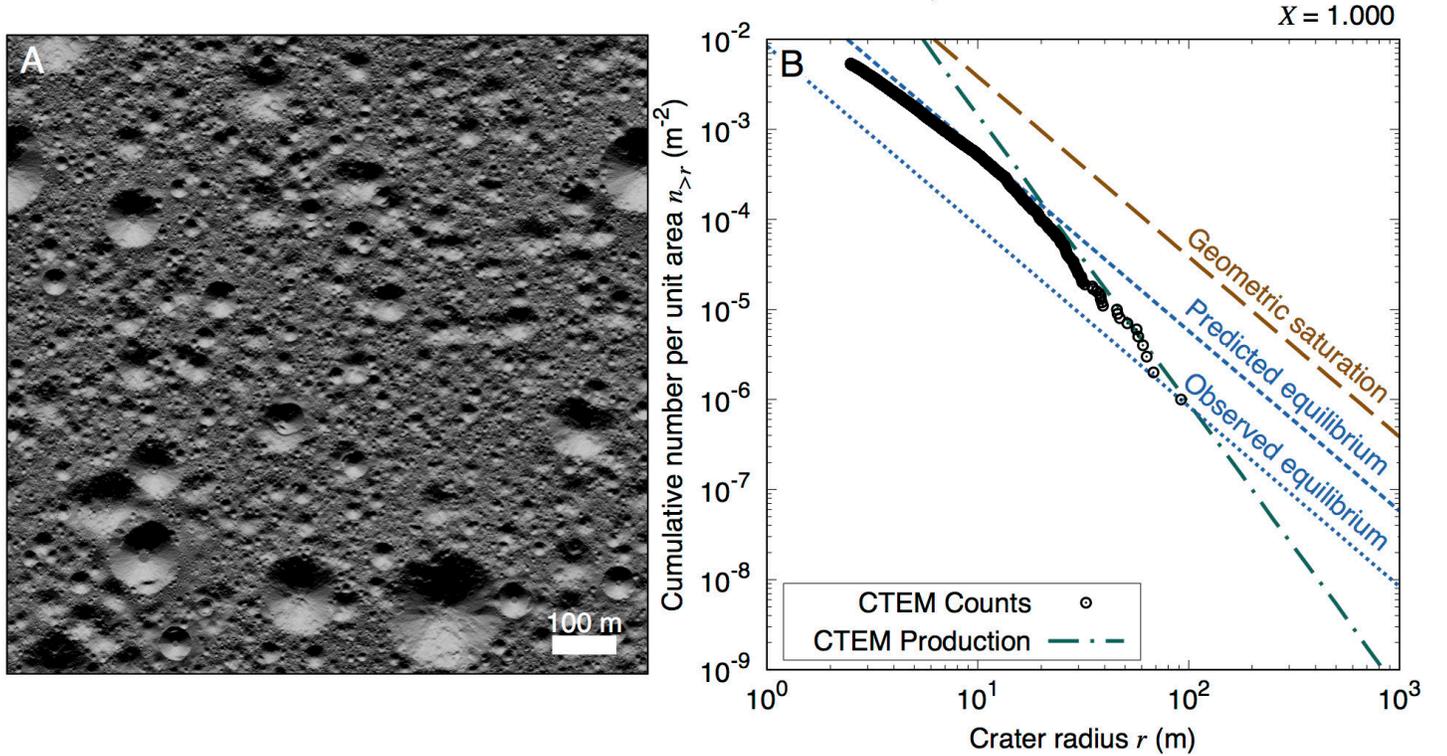

Figure 9. CTEM simulation at dimensionless time $X = 1$ of the slope-dependent proximal ejecta distribution model for diffusive degradation of small craters by large craters. In this model, we only consider primary impactors. A) The simulated appearance of the CTEM-generated DEM at a sun angle of 45° from the top of the image at B) The dash-dot line (teal) is the production function with parameters $n_{p,1} = 4.3$ m$^{1.2}$ and $\eta = 3.2$, for $r > 6$ μm craters. Circles show the cumulative SFD of the crater counts in the CTEM simulation. The line labeled "Observed equilibrium" is our Fit 1 SFD with $n_{eq,1} = 0.0084$ and $\beta = 2$. The line labeled "Predicted equilibrium" was using equation (30) using visibility function parameters $K_{v,1} = 0.17$ and $\gamma = 2$, and degradation parameters $K_{d,1} = 0.27$, $\psi = 2$, and $f_e = 1$.

Craddock and Howard (2000) assumed that the degradation of lunar craters in the size range of $500\text{ m} < r < 1500\text{ m}$ was driven by $\sim 1\text{ mm}$ micrometeoroids. Therefore, one possibility is that we are not modeling the micrometeoroid population correctly. Our results from Section 3.1 suggest that we need a higher amount of diffusive degradation in order to match the observed equilibrium density for small lunar craters. Therefore, in order for the Craddock and Howard assumption to be correct, in which micrometeoroids dominate the diffusive degradation of lunar landscapes at the scale of our study, the production SFD must have more micrometeoroids than is predicted by a simple power law extrapolation of the NPF to small sizes. We therefore consider the possibility that the production function is "enhanced" with an additional population of micrometeoroids. We will first constrain what such a population would look like in order to match the observed degradation state of the small craters of our comparison data set, and then we will compare this constraint with observations of what the true micrometeoroid population looks like.

We can use our analytical models developed in Section 2 to predict the effect of a large population of unresolved micrometeoroid craters, which will be smaller than the 1 m resolution limit of our CTEM simulations. We will assume that all craters (including our micrometeoroids) follow the uniform circular degradation function we derived from our single crater size degradation experiments in Section 3.1, where $K_{d,1} = 0.27\text{ m}^2$, $\psi = 2$, and $f_e = 1$.

Consider a steep-sloped production SFDs (i.e. $\eta > 2 + \psi$). The integral given by equation (27) predicts that $k' \to \infty$ as $\check{r} \to 0$. Physically this means the smallest craters in such a steep-sloped SFD would wipe the surface clean of all large craters faster than they could accumulate, and the surface would never be able to retain any countable craters. This is obviously not what happens in nature! Now instead consider a production function that has $\eta > 2 + \psi$, but only for a finite range of craters sizes, such that $\check{r} > r_{min} > 0$.

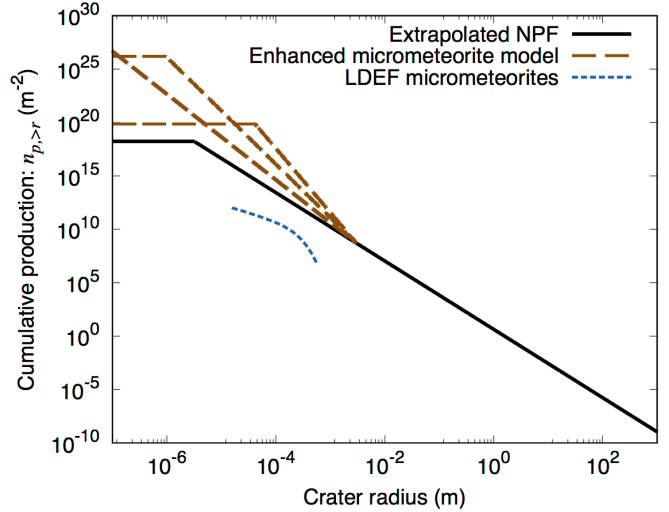

Figure 10. The solid line (black) is the production function produced by extrapolating the of the NPF, with a slope of $\eta = 3.2$, to craters as small as $r \sim 6\ \mu\text{m}$. The short dashed line (blue) is the production of micrometeoroid craters constrained from the LDEF experiment and iSALE simulations of Cremonese et al. (2012). The long-dashed lines (dark orange) show a set of enhanced micrometeoroid production functions. These each have a steep branch (shown are $\eta_{mm} = 4.01, 5,$ and 6) for $r < 0.03$ m that produces the equivalent of a constant degradation rate of $K' = 650\text{ m}^2$ in CTEM simulations with a resolution of 1 m/pix.

In such a broken power-law production function, degradation could again have a finite value, provided that the production function slope had a roll-over to $\eta < 2 + \psi$ below $r_{min}$.

We now construct an "enhanced micrometeoroid" production function for our CTEM simulations. We do this by creating a broken power-law production function that has three branches. The first branch is our "resolvable" crater branch, which is simply the same production SFD that we have used previously, but only for craters larger than the resolution limit of our CTEM simulations, which we set as $r_b$. The second branch is our "micrometeoroid" branch, which has a much steeper slope than our resolvable branch and connects with our resolvable branch. Because our micrometeoroids are constrained to produce a finite amount of diffusive degradation, we must roll over the micrometeoroid branch to a "craterless" branch at some size $r_a < r_b$. With this production function the resolved craters will degrade the surface just as in Section 3.1, but now we add a new source of diffusion arising from the unresolved micrometeoroids, which will have a steeper slope than the resolved craters. Because the micrometeoroids are unresolved, they will create the equivalent of a constant diffusivity on the resolved craters.

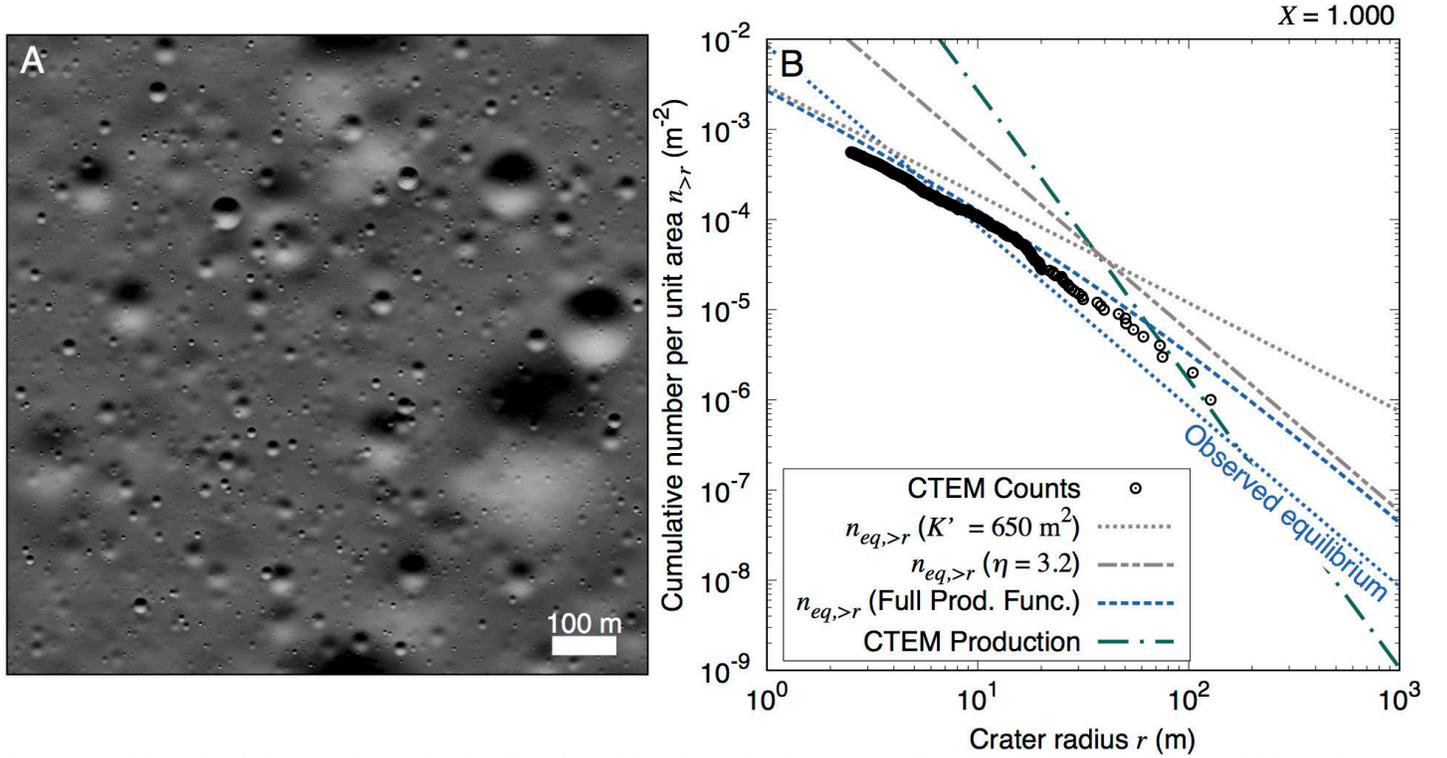

Figure 11. CTEM simulation for the production function with enhanced micrometeoroid population (see Figure 10) which produces an equivalent of a constant degradation rate of $K' = 650$ m$^2$ for the resolved craters ($r > 1$ m). The resolvable crater production function has parameters $n_{p,1} = 4.3$ m$^{1.2}$ and $\eta = 3.2$, and our visibility function has parameters $K_{v,1} = 0.17$ and $\gamma = 2$. We show three different solutions to the equilibrium SFD, $n_{eq,>r}$ here. The line labeled ($K' = 650$ m$^2$) is the constant degradation rate solution given by equation (18). The line labeled ($\eta = 3.2$) is the crater size-dependent degradation rate solution given by equation (30), using the resolvable crater production function. The line labeled (Full Prod. Func.) is a numerical solution to the equilibrium SFD using a piecewise degradation rate parameter given by equation (36). The CTEM crater counts match the predicted equilibrium line, but do not resemble the observed equilibrium SFD.

For the micrometeoroid branch where $r_a < r < r_b$. $\eta_{mm} \gg \eta$ we make use of the accumulation rate of the degradation state, $K'$, rather than the non-dimensional crater degradation parameter $k'$. As long as the cratering rate is constant in time then $K' = \kappa$, which is the more commonly-used diffusivity. The unresolved micrometeoroid craters will be fully in the partial degradation regime, so using equation (25) we can write

$$K'_{mm} = -\int_{r_a}^{r_b} K_{d,1} \pi f_e^2 n_{p,1} \eta \check{r}^{\psi - \eta + 1} d\check{r}. \quad (34)$$

Evaluating the integral given by equation (34):

$$K'_{mm} = K_{d,1} \pi f_e^2 n_{mm,p,1} \left( \frac{\eta_{mm}}{\psi - \eta_{mm} + 2} \right) \left( r_a^{\psi - \eta_{mm} + 2} - r_b^{\psi - \eta_{mm} + 2} \right), \quad (35)$$

Our CTEM resolution is 1 m/pix, but CTEM can model the ejecta blankets of craters below the resolution limit. In order to ensure that our enhanced micrometeoroid population behaves in a way that is consistent with a constant diffusive degradation rate, we chose a value of $r_b = 0.03$ m , which is comfortably below the resolution limit. So to create our micrometeoroid population, we create a steep branch of our production function that has $\eta_{mm} > 2 + \psi$ for $r_b < 0.03$ m. For $r > 0.03$ m we follow our original extrapolated NPF, with parameters $\eta = 3.2$ and $n_{p,1} = 4.3$ m$^{1.2}$. We also require that the broken power law is continuous at the break at $r_b = 0.03$ m.

We constrain the $r_a$ and $\eta_{mm}$ of our micrometeoroid model such that degradation state generated by our micrometeoroids will be $K'_{mm} = 650 \text{ m}^2$. This value was chosen so to approximately match what is required for the $r \sim 1 - 10$ m craters to be in equilibrium. Using the same degradation function that we derived from our single crater size experiments in Section 3.1 ($K_{d,1} = 0.27$, $\psi = 2$, and $f_e = 1$), we can use equation (35) to generate a set of enhanced micrometeoroid production function models that result in the same value of $K'_{mm}$. We plot examples of these production functions compared to our extrapolated NPF in Figure 10.

The analytical model developed in Section 2.2 predicts that a constant diffusivity model would result in an equilibrium SFD for the resolved craters with a slope $\beta = \eta - \gamma$, given by equation (18). For our resolved craters, $\eta = 3.2$ and $\gamma = 2$, and so we would predict an equilibrium slope of $\beta = 1.2$, rather than the observed value of $\beta \sim 2$. As discussed in Section 0, an equilibrium slope of $\beta = 2$ implies both geometric similarity of the visibility and degradation functions and that the countable craters originate from a population with the same slope that dominates the degradation. While we have imposed geometric similarity on these CTEM simulations ($\gamma = 2$, $\psi = 2$, and $f_e = 1$), the diffusive degradation of the surface is dominated by the production of our enhanced micrometeoroid population, which has a steeper slope than that of our resolved craters, and this should drive the equilibrium slope away from 2. This suggests that micrometeoroids are unlikely to be the extra source of diffusive degradation required to match the observed empirical equilibrium SFD.

To compare with our numerical model, we note that our equilibrium SFD contains both the effects of micrometeoroids and the resolvable craters. Therefore the analytical model for the equilibrium SFD for constant degradation rate, $K'$, given by equation (18) is only an approximation. To account for the contribution by both the micrometeoroid branch of the production function and that from the resolvable branch of the production function, we must modify our dimensonless degradation parameter such that:

$$k' = k'_{mm} + k'_{res}, \tag{36}$$

Where:

$$k'_{mm} = \frac{1}{K_{v,1} r^\gamma} \int_{r_a}^{r_b} K_{d,1} \pi f_e^2 n_{mm,p,1} \eta_{mm} \check{r}^{\psi - \eta_{mm} + 1} d\check{r}, \tag{37}$$

and

$$k'_{res} = \frac{1}{K_{v,1} r^\gamma} \int_{r_b}^{r_{crit}} K_{d,1} \pi f_e^2 n_{p,1} \eta \check{r}^{\psi - \eta + 1} d\check{r} + \int_{r_{crit}}^{\infty} \pi f_e^2 n_{p,1} \eta \check{r}^{-\eta + 1} d\check{r}, \tag{38}$$

The resulting equilibrium SFD is determined by substituting the above expressions for $k'$ into equation (17). This will not result in a power law equilibrium SFD. However, it should asymptotically approach the SFD given by equation (18) for $r < r_b$, and likewise should asymptotically approach the SFD given by equation (30) when $r > r_b$.

As in the simulation in Section 3.1 we simulated the cratering due to our enhanced micrometeoroid population in CTEM with a resolution of 1 m/pixel on a domain $1000 \times 1000$ pixels. The final result of this simulation is shown in Figure 11. As predicted, the small craters of our simulation of the enhanced micrometeoroid production population do not produce a power law equilibrium SFD, and instead it approaches the equilibrium SFD predicted by our constant degradation rate model given by equation (18) with slope of $\beta = 1.2$ at small crater sizes, and at large craters the equilibrium SFD approaches tthe same equilibrium SFD as we obtained in the simulation shown in Figure 9 with a $\beta = 2$ but a coefficient an order of magnitude higher than is observed.

We can also compare the population of enhanced micrometeoroid craters to constraints on the observed flux of micrometeoroids. Cremonese et al. (2012) used the iSALE hydocode to model the micrometeoroid impacts accumulated on Long Duration Experimental Facility (LDEF). We plot the cumulative production SFD for the LDEF-derived micrometeoroid craters in Figure 10. These results show that even our simple extrapolation of the NPF to small sizes produces orders of magnitude more micrometeoroid impacts than is observed, and yet still falls short of producing the diffusive degradation required to match equilibrium. Both the observational constraints on the flux of micrometeoroids and the slope of the observed equilibrium SFD suggest that the additional diffusive degradation required to produce crater count equilibrium of simple lunar craters of the maria is not due to micrometeoroids.

### 3.3. Modeling the equilibrium SFD for the spatially uniform crater size-dependent degradation region model in CTEM.

In Sections 3.1 and 3.2 we used numerical simulations in CTEM to show that neither the slope-dependent proximal ejecta distribution of primary impactors, nor an enhanced micrometeoroid population can produce the right kind of diffusive degradation required to match the observed equilibrium SFD seen in small simple craters of the lunar mare, as shown in Figure 2. Under the assumption of geometric similarity ($\gamma = \psi = 2$), the crater size-dependent degradation model predicts the correct slope of $\beta = 2$ for the equilibrium SFD, regardless of the production function slope, $\eta$, as expressed by equation (31). These results suggest that in order to match the observed equilibrium SFD of small lunar craters, we require some source of extra diffusion that is crater size-dependent. We therefore turn the problem around and use the equilibrium SFD parameters of $n_{eq,1} = 0.0084 \approx 0.02 n_{gsat,1}$ and $\beta = 2$ as a constraint on the parameters of the degradation function, using our analytical model given in equation (30). Our goal is to quantify the required "extra degradation" that each new crater contributes during its formation. For now, we will also make the same simplifying assumption as we did in developing our analytical model in Section 0 that the degradation region is uniform and circular. We will later relax this assumption when we explore the effect of spatial heterogeneity in the form of distal rays in Section 3.4.

In the crater size-dependent degradation model, each new crater contributes to the degradation of the pre-existing landscape in an amount that is proportional to its size via the degradation function. The degradation function acts over an extended region (see Figure 4A). The source of degradation could come from a number of physical processes, such as seismic shaking, secondary craters, ballistic sedimentation, etc.

Before we proceed, we briefly mention that we also found via numerical experimentation that there is an upper limit to how high $K_d$ can be for a given $f_e$. When $K_d$ gets high enough, the extra per crater degradation can become powerful enough to completely wipe clear any pre-existing topography. Once this happens, additional degradation no longer contributes to the actual degradation state of the surface. We call this effect *the diffusion limit*. This is somewhat similar to the $I = 0$ case of Howard (2007), in which no memory of the local pre-existing terrain remains after each cratering event. However, craters on real planetary surfaces inherit some memory of the pre-existing state in both the morphology of the crater interior and proximal ejecta on slopes (Aschauer and Kenkmann, 2017; Howard, 2007) and the visibility of pre-existing craters beneath ejecta blankets (Fassett et al., 2011). Therefore, $K_d$ must be well below the value of the diffusion limit to be consistent with observations. With some numerical experiments in CTEM we found that that $K_{d,1}\big|_{limit} \approx 0.76 f_e^{1.5} \ \text{m}^2$. This sets an upper bound on what our per-crater degradation function coefficient, $K_{d,1}$, can be in order for the model to reproduce the correct equilibrium SFD coefficient, $n_{eq,1}$.

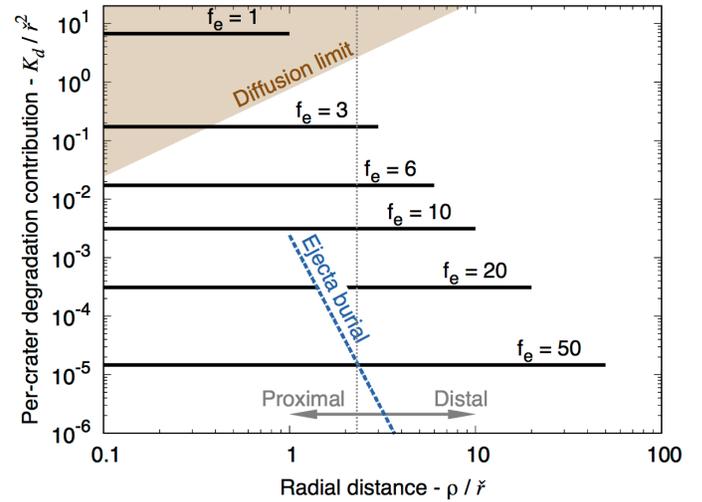

Figure 12. This plot shows the contribution to the diffusive degradation state, $K_d$, by the formation of each new crater of radius $\check{r}$ as a function of radial distance from the crater center, $\rho$. All values have been normalized by the crater radius under the assumption of geometric similarity ($\gamma = \psi = 2$). Each solid black line shows a solution for the equilibrium SFD ($n_{eq,1} = 0.0084$ and $\beta = 2$) for the crater size-dependent degradation model, given by equation (32) assuming a uniform circular degradation region with radius $\check{r} f_e$. The shaded region (orange) shows where values of $K_d / \check{r}^2$ that are not attainable for a given value of $f_e$ (the diffusion limit), which rules out the $f_e = 1$ solution. The dashed line (blue) shows the effectiveness of ejecta burial, given by equation (33) for an ejecta blanket profile $h_{ej} / \check{r} = 0.04 (\check{r}/\rho)^{-3}$ (McGetchin et al., 1973). Only the solution for $f_e = 50$ satisfies the constraint that the extra degradation is less effective than ejecta burial in the proximal ejecta blanket region.

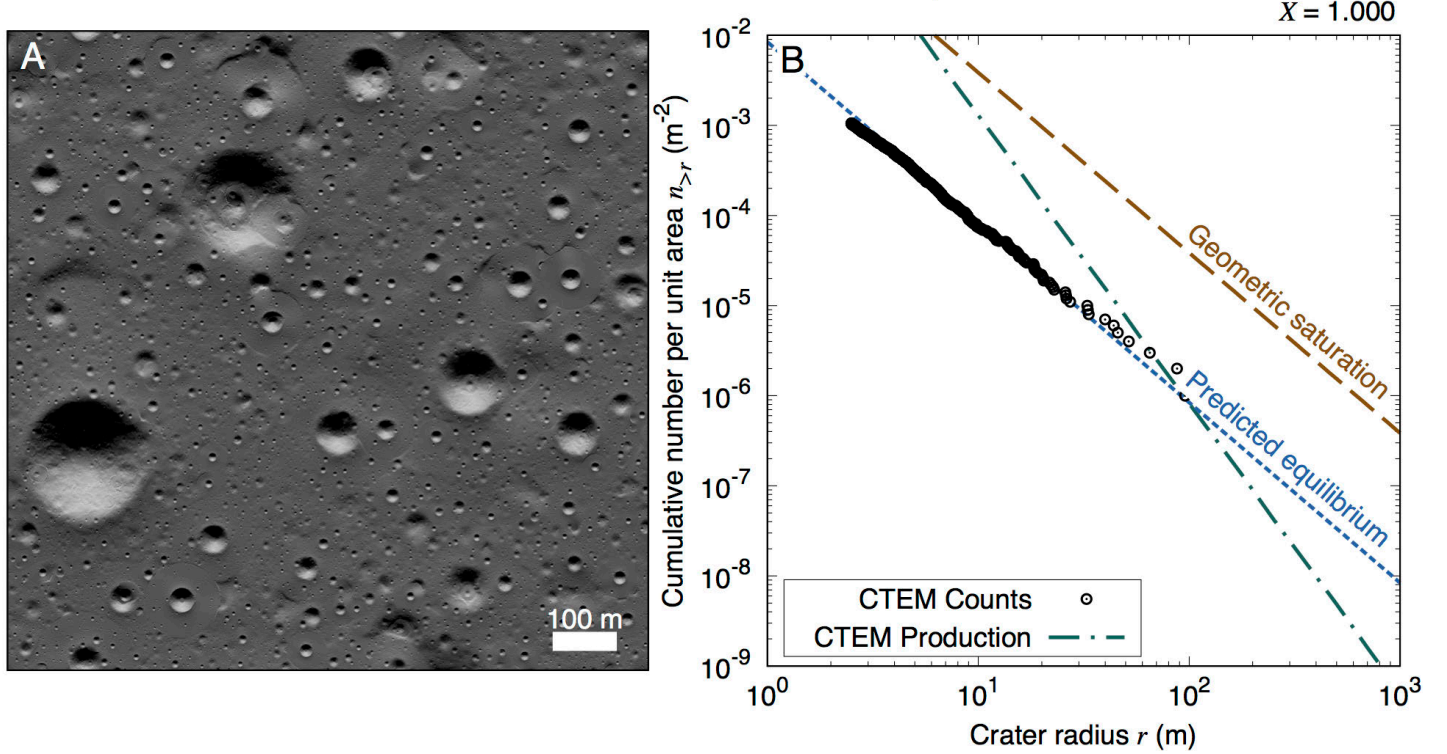

Figure 13. Similar to Figure 9 but with additional extra diffusion added over a uniform region with radius $f_e r$, with $f_e = 3$ and $K_{d,1}$ determined by solving equation (32). While we are able to reproduce the equilibrium SFD correctly, the model results in a surface that does not resemble the lunar surface. Each new crater produces a smooth crater free "halo" out to the radius of $rf_e$, which is not observed on natural surfaces (see Figure 2A).

We impose geometric similarity, which constrains our degradation function slope to $\psi = 2$ (as well as the visibility function slope of $\gamma = 2$). Under these model assumptions, two parameters remain unconstrained: the degradation state coefficient $K_{d,1}$, and the degradation size scale factor, $f_e$. These are anti-correlated, such that in order to achieve a given value of $n_{eq,1}$, a larger value of $K_{d,1}$ is needed for a smaller value of $f_e$, and vice versa. Figure 12 shows a family of solutions for the combination of $K_d$ and $f_e$ required to match the equilibrium SFD coefficient $n_{eq,1} = 0.0084$. For reference we also plot both the diffusive limit region and our model for the degradation strength of ejecta burial from equation (33).

There are a number of constraints that can determined from Figure 12. First, this figure shows that the solution for $f_e = 1$ violates the diffusion limit constraint. This means that the extra degradation we require involves processes outside the crater rim. Figure 12 can also be used to show that ejecta burial is unlikely to be an important process for setting the equilibrium SFD. In the proximal ejecta region the effectiveness of ejecta burial is several orders of magnitude lower than the equilibrium solution for the equivalent degradation region ($f_e = 3$). The effectiveness of ejecta burial decays rapidly with distance from the crater rim, so simple burial by distal ejecta (i.e. low energy deposition) is not likely to play much of a role in crater degradation, in contradiction to the conclusion by Marcus (1970) that burial by distal ejecta was the dominant degradation mode for small lunar craters.

The comparison between the effectiveness of ejecta burial and solutions to the degradation function shown in Figure 12 can also be used as a constraint in a different way. As we discussed when developing our model for the effectiveness of ejecta burial, Fassett et al. (2011) used crater counts surrounding Orientale and a simple ejecta burial model to show that the Orientale ejecta thickness profile was consistent with estimates of McGetchin et al. (1973). This can only have been done if simple ejecta burial (i.e. low energy ejecta deposition) is the dominant mechanism for degrading and removing pre-existing craters in Orientale's ejecta blanket. The implication is that the extra degradation we require to match the observed value of $n_{eq,1}$ cannot be stronger than ejecta burial in the proximal ejecta region, or this observational constraint would be violated. This is not a very strong constraint, as it could be that the ejecta of large basins such as Orientale are less effective at degradation, relative to their size, than small simple craters. Nevertheless, it serves as a useful limit to compare the relative strength of our "extra" degradation in the proximal ejecta region.

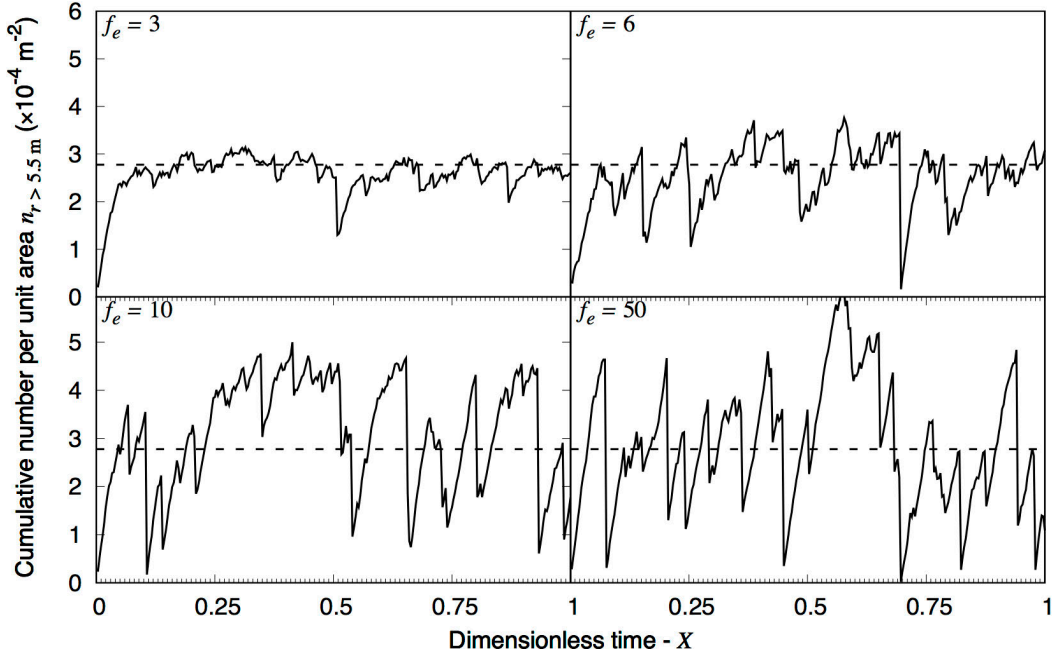

Figure 14. Cumulative surface density of craters with $r > 5.5$ m from our CTEM simulations with extra diffusive degradation added in order to match the observed equilibrium SFD with parameters $n_{eq,1} = 0.0084$ and $\beta = 2$, given by equation (32). The extra diffusive degradation generated by each crater of radius $\check{r}$ is modeled using a uniform circular region with degradation function coefficient $K_{d,1}$ over a region of radius $\check{r} f_e$. The dashed line indicates the observed equilibrium crater density $n_{eq,>5.5\,m}$. In the uniform degradation region model, the number density of the simulated craters fluctuates around the equilibrium value. As the value of $f_e$ increases from 3 to 50, the magnitude of the fluctuations increases.

Using the effectiveness of ejecta burial in the proximal ejecta region as an upper limit of the extra degradation required for setting the small crater equilibrium SFD, Figure 12 shows that the only solutions that work are those with $f_e \gtrsim 50$. This upper limit suggests that the required extra degradation takes the form of energetic distal ejecta deposition, which includes secondary cratering and ballistic sedimentation (see Section 1.4.4). To determine what kind of degradation function best matches observations, we conducted a series of numerical experiments in CTEM in which we add extra diffusive degradation to each crater over a circle of radius $f_e \check{r}$ for $f_e \geq 3$. We assume a geometrically similar form for the degradation function ($\psi = 2$ and constant $f_e$) and calculate the value of $K_{d,1}$ required to match our Fit 1 equilibrium cumulative SFD for a given value of $f_e$. Our input production SFD is a simple power law function with a slope $\eta = 3.2$, and we include the diffusive effects of subpixel craters down to $\sim 6\ \mu m$.

In previous CTEM simulations, the repeating boundary condition was adequate for modeling local scale effects of craters on our domain. However, once we consider that the distal effects of craters, we need to consider the effects of large craters that may form beyond our simulated local domain. We implemented a *superdomain* in CTEM that is large enough to accommodate the largest craters that could affect our local domain for a given value of $f_e$. Just as in our local domain, superdomain craters are randomly drawn from our production function. If a superdomain crater is found to be large enough to affect the local domain, then distal diffusive degradation is applied to the local simulated domain.

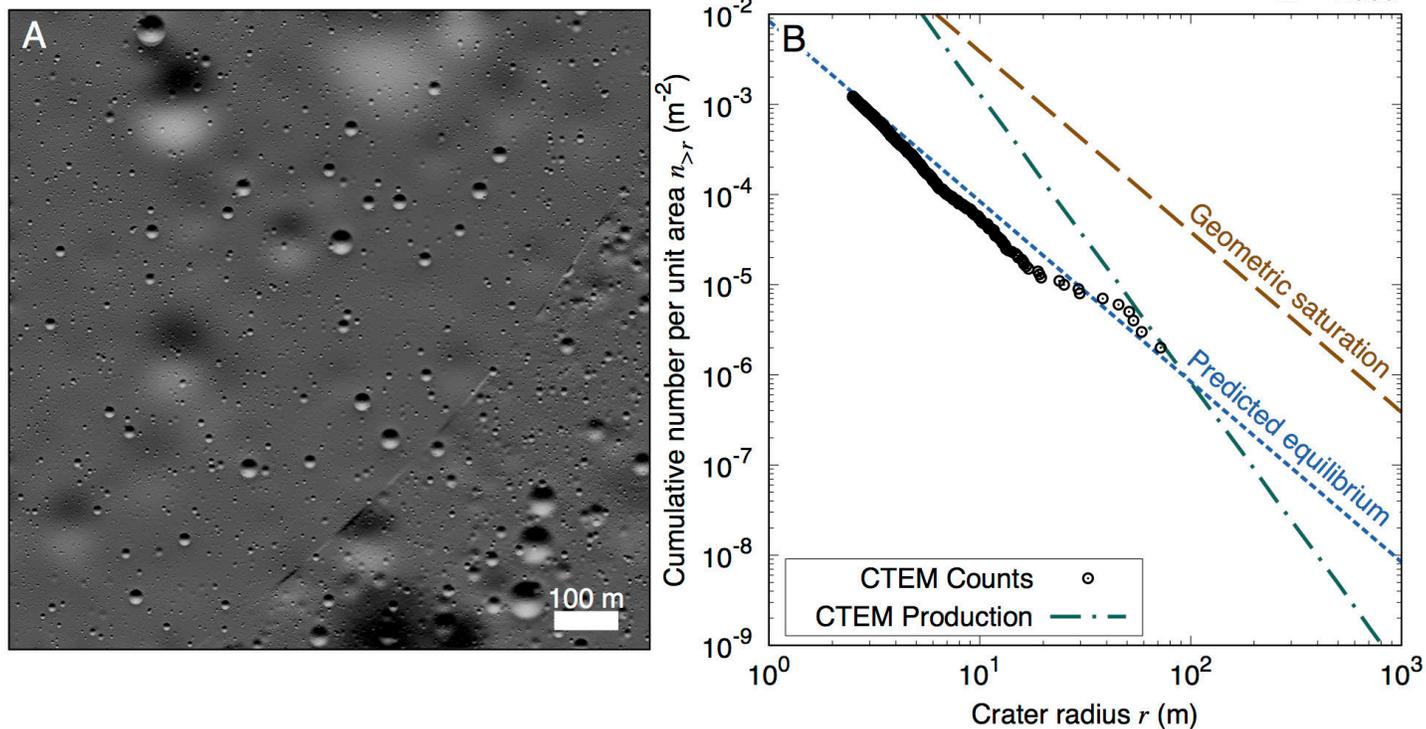

Figure 15. Similar to Figure 13 but with $f_e = 10$. A larger value of $f_e$ decreases the magnitude of the "halo" effect, but the increases the frequency of anomalous jumps from distant craters.

Figure 14 shows the cumulative number of craters with $r > 5.5$ m vs. simulated time for four of our CTEM simulations. We also show the predicted cumulative number of craters in equilibrium from equation (30).These simulation results show that for large values of $f_e$ we start to see large variations in crater density. They are caused by large distant craters from the superdomain that occasionally obliterate craters on the local simulation domain. This type of behavior is typical of anomalous diffusion, rather than classical diffusion, and we call them "anomalous jumps."

It would appear from Figure 14 that our $f_e = 3$ case would provide the closest match to the observed lunar surface, as no large anomalous jumps occur within the simulation time. However, even though the crater counts are well-matched in the $f_e = 3$ case, the surface morphology tells a different story.

Figure 13A shows a shaded DEM of the surface. Each crater in this simulation produces a smooth halo of obliterated craters, which is not observed in real craters. This is because the per-crater degradation function for this case is very close to diffusive saturation, as seen in Figure 12.

Increasing $f_e$ to 10 reduces the effect of diffusive saturation (Figure 15) but begins to introduce anomalous jumps that periodically obliterate the craters in the simulated surface. As $f_e$ increases to 50, these anomalous jumps become more frequent, which likely violates observational constraints. That is, if these anomalous jumps occurred as often as these high $f_e$ value simulations would suggest, a significant fraction of the maria would smooth and nearly crater free at the scales comparable to our simulation domain.

### 3.4. Modeling the equilibrium SFD for the spatially heterogeneous distal degradation model in CTEM.

In Section 3.3 we defined a model in which each new crater of size $\check{r}$ contributes to the diffusive degradation of the pre-existing surface over a uniform circular region of size $f_e\check{r}$. We showed that the value of the degradation region scaling factor $f_e$ was constrained to be quite large ($f_e \gtrsim 50$) in order to not violate observations of proximal ejecta burial. However we showed that uniform circular degradation functions with large values of $f_e$ produce too-frequent anomalous jumps, which also violate observational constraints from the lunar surface. On the face of it, these results suggest that there are no solutions to the problem of matching empirical equilibrium. In order to solve this problem, we need to understand the physical meaning of large values of $f_e$.

When $f_e \gtrsim 2.3$, then we are assuming that each new crater is producing diffusive degradation of terrains in the region of its distal ejecta. Thus far when we have modeled this additional distal degradation in CTEM we have made the simplifying assumption that the per-crater degradation $K_d$ was constant over a circular region of radius $f_e\check{r}$. However, we know from observations that the ejecta blanket of craters becomes highly spatially heterogeneous beyond $\sim2.3\check{r}$, and the distal ejecta is distinguished by thin patchy streamers known as rays (Melosh, 1989).

It is reasonable to assume that the spatial distribution of diffusive degradation from a crater would correlate with its rays, rather than occurring over a uniform circular region. If so, then the contribution from distant large superdomain craters that causes the frequent anomalous jumps would be reduced, as the probability of a ray from a superdomain crater intersecting the domain would be lower than in the uniform circular degradation model of the same size degradation region.

To model the effect of crater rays, we apply our extra diffusive degradation in a ray pattern. We use ray geometry model modified from that described in Huang et al. (2017). That work reported a polar function that defined the ray boundaries. Here we use a similar formula that defines a spatial field function that is compatible with our formulation of the degradation function in Section 0. Our ray function is not a well-constrained depiction of the spatial variability of rays, as defining such a function would be beyond the scope of this work. Instead, we develop a function that captures some of the qualitative properties of crater rays (see Elliott et al., 2018) in order to explore, qualitatively, the effect of ray geometry on our CTEM simulation results.

In our model, the number of rays is given as $N_{rays}$. The strength of the degradation function is described using a spatial intensity function $f(\xi, \phi)$, where $\xi = \rho/\check{r}$. We model the intensity of the degradation function to follow a gaussian function across the rays, each of which is defined over a sector centered at azimuth angle $\phi_{r,i}$, where $i$ is the index of the individual rays. Each ray center is evenly distributed such that $\phi_{r,i} = 2\pi i/N_{rays}$. The relative length of each individual ray is given as $L_i$. The sectors are evenly divided up by the number or rays, however we randomize which sector is assigned which particular ray length. The intensity function is defined as:

$$
\begin{aligned}
&f(\xi, \phi) \\
&= K_{ray} \sum_i^{N_{rays}} \begin{cases} f_i \exp\left[-\dfrac{(\phi - \phi_{r,i})^2}{2[\phi_w(\xi)]^2}\right], & \xi < L_i \\ 0, & \xi \geq L_i \end{cases}
\end{aligned} \quad (39)
$$

where $K_{ray}$ is a constant scaling factor that is constrained by comparing the integrated intensity function to a uniform circular degradation. The length of each ray is controlled by the length parameter:

$$
L_i = L_1 \exp\left[\log\left(\frac{L_{N_{rays}}}{L_1}\right)\frac{(N_{rays} - i + 1)^2 - 1}{N_{rays}^2 - 1}\right]. \quad (40)
$$

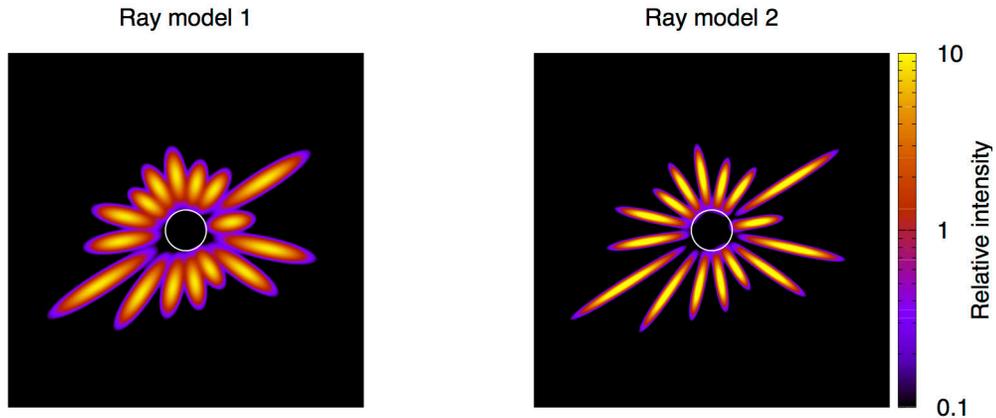

Figure 16. Two degradation intensity functions used to model the spatial heterogeneity of rays from equations (39)-(42). Both ray models use identical parameters with $N_{rays} = 16$, $L_1 = 3$ (the smallest ray length) and $L_{16} = 16$ (the longest ray length). Ray model 1 uses width parameter $w = 1$ and ray model 2 uses width parameter $w = 2$, which results in narrower rays.

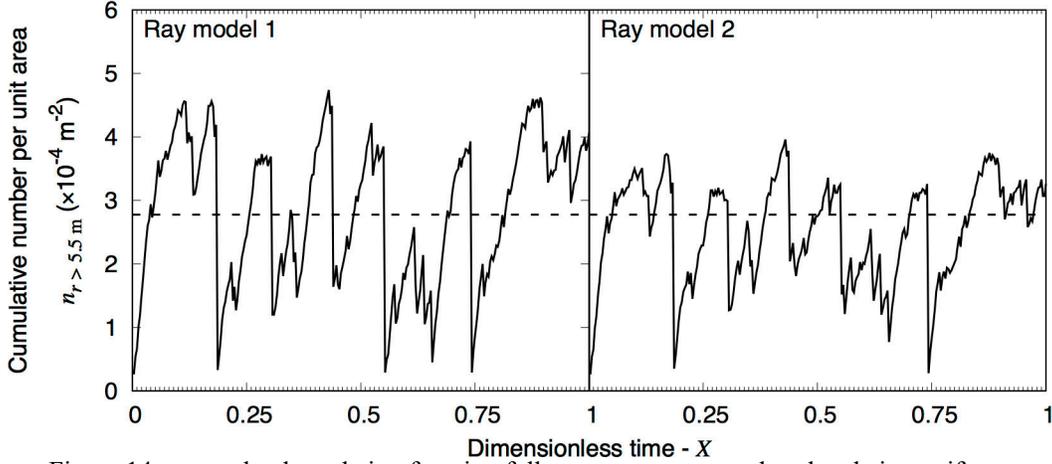

Figure 17. Similar to Figure 14 except the degradation function follows ray patterns, rather than being uniform over a circular region. Ray model 2 has narrower rays than ray model 1.

The width of each ray is controlled by the parameter $\phi_w(\xi) = r_w(\xi)/\xi$:

$$
\begin{aligned}
r_w(\xi) = \frac{\pi}{wN_{rays}} r_{cont} \Big\{ 1 \\
- \left(1 - \frac{2w}{r_{cont}}\right)\exp\left[1 - \left(\frac{\xi}{r_{cont}}\right)^2\right] \Big\},
\end{aligned} \quad (41)
$$

where $r_{cont} = 2.3$ is the continuous/distal ejecta boundary, and $w$ is a width scaling factor that we use to determine the relative width of the rays in two different models. We will call $w = 1$ "ray model 1" and $w = 2$ ray model 2. Finally, we radial dependence on the strength of the intensity is determined by:

$$
f_i(\xi) = \left(\frac{\xi - 1}{r_{p,i}}\right)^4 \exp\left[\frac{1 - \left(\frac{\xi - 1}{r_{p,i}}\right)^4}{4}\right]. \quad (42)
$$

where $r_{p,i} = \frac{1}{2}(L_i - 1)$ . This function causes the degradation function to have peak intensity at a point half way between the crater rim and the length of the ray, $L_i$. We generate the function this way so that the degradation function meets the observational constraint that the extra diffusion generated by this function cannot exceed that arising from our ejecta burial degradation model (see Figure 12), but it could be larger in narrow regions far from the crater rim.

We show two different ray patterns generated by our intensity function in Figure 16, using parameters $N_{rays} = 16$ , $L_1 = 3$ (the smallest ray length), and $L_{16} = 16$ (the longest ray length). Ray model 1 uses width parameter $w = 1$ and ray model 2 uses width parameter $w = 2$, which results in narrower rays. For each crater, we apply diffusive degradation only to regions within the ray. We also modified the superdomain craters to have rays.

We show the time evolution of countable crater numbers in Figure 17, which is similar to Figure 14, but with our two ray models. In both cases the magnitude of the anomalous jumps has reduced, and narrower rays of Model 2 generate less variation in crater density than the wider rays of Model 1. We show the output of a CTEM simulation using our narrow ray Model 2 in Figure 18. Both our ray geometry model and our constant distal degradation model are highly simplified, and our runs still show periodic large anomalous jumps. An improved model based on degradation in rays requires better constraints on the size-frequency distribution of ejecta fragments in distal ray ejecta, as well as the spatial distribution of ejecta fragments. Such modeling is beyond the scope of the present work. However even our simplified model presented here provides unique new constraints on distal ejecta degradation.

## 4. Discussion and Conclusions

In this paper we used the empirically-defined crater count equilibrium cumulative size-frequency distribution as a constraint on diffusive topographic degradation of the lunar surface. We derived a new diffusion-based analytical model that quantifies the equilibrium crater count cumulative size-frequency distribution, given in equation (30). We also performed a simulations of cratered surfaces using the Monte Carlo landscape evolution code, CTEM. An important outcome of this combined analytical and numerical modeling approach is that the numerical model can be used to test the robustness of the analytical model. With the exception of the subpixel resolution craters, the CTEM simulations model the formation of each individual crater, with all of the associated degradation processes. Therefore, CTM directly simulates much of the complexity of real cratered landscapes, which are averaged out in the analytical model.

We showed in Sections 3.3 and 3.4 that one of these assumptions, that the diffusive distal degradation region could be approximated as its spatially uniform average, lead to differences in the behavior of the numerical model compared with the analytical model. The anomalous jumps seen in the simulations with large values of the degradation region scale factor, $f_e$, were less apparent in the $f_e = 3$ cases, even if the smoothed halos of the proximal ejecta region of each crater did not appear to match observations. Nevertheless, the lack of anomalous jumps in the $f_e = 3$ allowed us to test the robustness of the analytical model in predicting the correct equilibrium SFD, as given by equation (30). To this end, we performed a

**Crater size-dependent degradation (Ray model 2):** $K_{d,1} = 3.127 \times 10^{-3}$; $f_e = 10$; $\psi = 2.0$

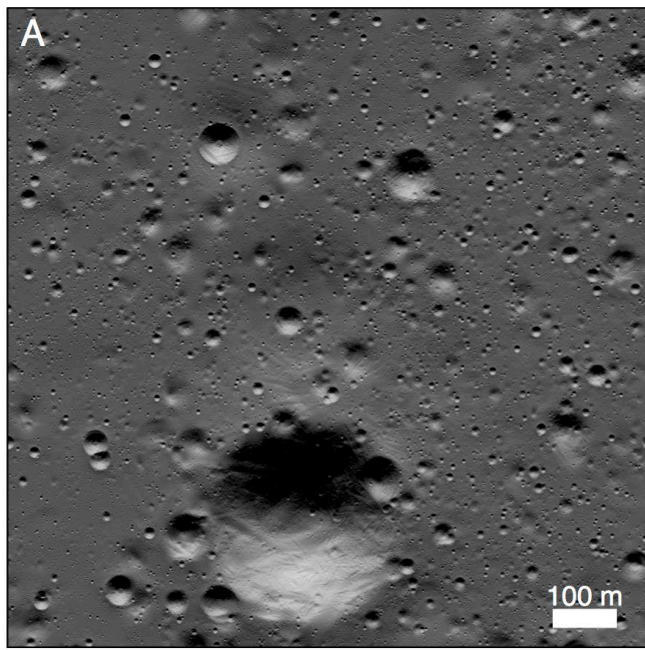

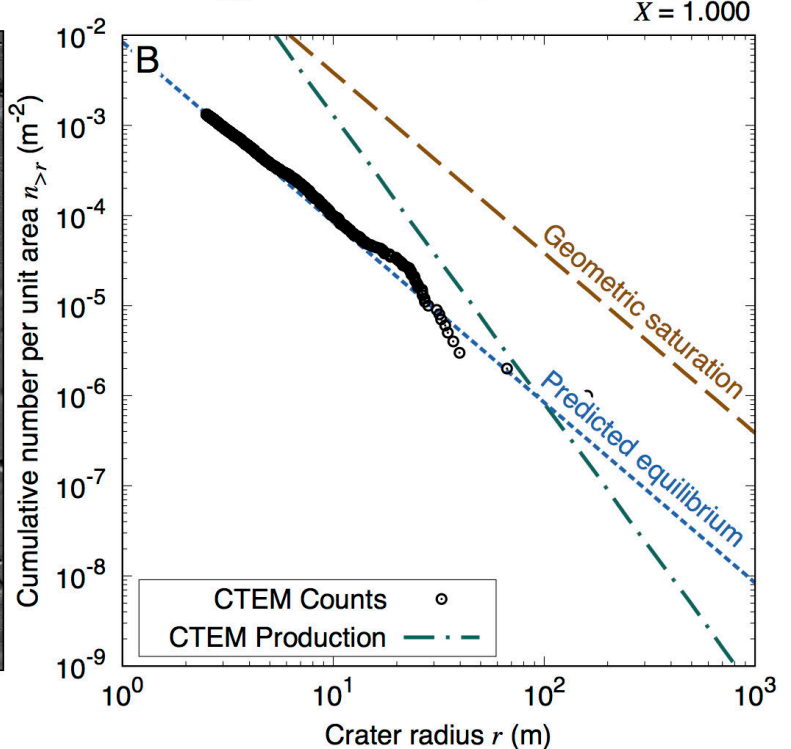

Figure 18. This is a frame of a simulation with $f_e = 10$ but with the degradation applied over a region occupied by spatially heterogeneous rays using ray model 2 shown in Figure 16. The low probability of a ray from a distant large crater intersecting the simulation region reduces the occurrence of large anomalous jumps.

suite of test simulations in which we varied the production function slope, $\eta$, and the degradation function slope, $\psi$. In all cases, the analytical model for the predicted equilibrium SFD correctly predicted the numerically-determined equilibrium SFD.

The constraints on our model for small crater equilibrium are the observed equilibrium SFD of crater counts in the lunar maria for craters with $r \lesssim 100$ m, the crater production function, and a visibility function that can be constrained from human crater count experiments. Our model also contains a degradation function that quantifies how much each new crater contributes to the diffusive degradation of the lunar surface. We discussed the development of our model and these constraints in Section 2. Using observations of crater counts in equilibrium and constraints from the morphology of lunar surfaces, we derived constraints on the degradation function in Section 3.

In Section 3.1 we tested a model similar to one developed by Soderblom (1970) in which the majority of diffusive degradation caused by crater formation is due to excavation and preferential downslope ejecta deposition by primary impactors. Under the assumption that the production function slope of $\eta = 3.2$ could be extrapolated down to the micrometeoroid sizes of $r = 6$ μm craters, we found that our predicted equilibrium SFD had crater number density nearly an order of magnitude higher than is observed. Furthermore, due to a phenomenon we call diffusive saturation it is impossible for any model of crater degradation that is restricted to the proximal regime to reach crater densities as low as the observed empirical equilibrium value.

In Section 3.2 we explored the hypothesis that an enhanced micrometeoroid population could be responsible for generating the extra diffusive degradation required to match the observed equilibrium value. This hypothesis fails for two different reasons. First, the required population of enhanced micrometeoroids is many orders of magnitude more than what is constrained from observations (see Figure 10). Second, the enhanced micrometeoroid population creates a non-power law equilibrium SFD that has a shallow slope of $\beta = 1.2$ for the small craters, and transitions to a steep slope of $\beta = 2$ at the larger craters, but with the an equilibrium coefficient that is an order of magnitude too high. Such a result is at odds with observations of the small crater equilibrium SFD across the lunar surface (e.g. Xiao and Werner, 2015). Therefore we rule micrometeoroids as an important process driving the diffusive degradation of lunar landscapes at the meter scale and larger.

We showed in Section 0 that an equilibrium SFD slope of $\beta = 2$ can occur for a crater size-dependent diffusive degradation model. Such a model requires that the source population that generates diffusive degradation originates in the same population of craters that is being counted. In Section 3.3 we performed simulations in which each crater generated a uniform region of diffusive degradation that scaled with crater radius, and used empirical equilibrium as a constraint on what we term the degradation function. We showed that the anomalous jumps in crater density occur more frequently as the size of the degradation region increases, but observations of crater removal by proximal ejecta burial constrain our degradation function to have a large radius, with $f_e \gtrsim 50$ (see Figure 12). The constraints that drive the degradation region to be large are therefore in opposition to the constraints imposed by the anomalous jumps in crater density. These contradictions can only be resolved if we consider that the distal degradation region is a uniform circular region, but is controlled by the ray pattern seen in relatively fresh craters.

In Section 3.4 we showed that if the degradation function had a spatially heterogeneous intensity similar to a ray pattern seen in distal ejecta, then the problem of frequent anomalous jumps could be suppressed. The heterogeneous nature of rays has also been shown to be important for lunar regolith compositional evolution (Huang et al., 2017), and the underlying energetic deposition that that creates rays likely extends much farther than they appear (Elliott et al., 2018).

Our major results suggest that crater equilibrium is controlled primarily by the highly energetic collision of distal ejecta fragments onto the lunar surface. This is perhaps a rather surprising and non-intuitive result, as the most prominent visible evidence for an impact event is in the formation of the primary crater and its proximal ejecta. The amount of topographic degradation caused by the distal ejecta appears relatively small in comparison. By relatively small, we mean that the quantity of degradation at any point in the distal degradation region is orders of magnitude smaller than the amount of degradation caused by the direct excavation of the crater, its deposition in the proximal ejecta, and removal of craters by burial under proximal ejecta (see Figure 12).

However, because the distal degradation occurs over a vastly larger area than the proximal degradation, this relatively small amount of distal degradation dominates the topographic evolution of lunar surface features and is primarily responsible for setting the equilibrium size-frequency distribution. This implication is consistent with recent observations of rapid regolith overturn generated by distal secondaries from recent craters (Speyerer et al., 2016).

Our distal degradation model contains a number of simplifying assumptions, but constraining all of them further is beyond the scope of the present work. For instance, we generated crater ray patterns that were geometrically similar. However, observations of the lengths of rays suggest that distal ejecta does scale with dimension (Elliott et al., 2018). Yet even if the degradation and/or visibility functions have some scale dependence, our model predicts that the equilibrium SFD will still have $\beta \sim 2$, and therefore small deviations in the observed equilibrium SFD away from the $\beta = 2$ could be indications of scale dependence in the real process of how craters form and affect the pre-existing terrain. A better degradation function could be derived from either a model for diffusive degradation by ballistic sedimentation, or observations of diffusive degradation within distal ejecta deposits. Such modeling would require better constraints on the spatial and size-frequency distributions of distal ejecta fragments, which are poorly understood.

Due to our results that energetic deposition of distal ejecta plays a critical role in setting the small crater equilibrium SFD, differences in the behavior of ejecta on different kinds of planetary terrains could results in differences in the equilibrium SFD. For instance, observations suggest that craters on Mercury produce proportionally more secondaries than those of the Moon (Strom et al., 2008). This could potentially explain why rates of diffusive degradation also appear to be faster on Mercury (Fassett et al., 2017). Bierhaus et al. (2018) showed that there is a great deal of variation in the production of secondary craters on bodies across the solar system, which are influenced by target material properties, the surface gravity, and escape velocity of each body. Also, while we have shown that the equilibrium slope is only weakly dependent on the slope of the production function, the coefficient does depend on equilibrium slope. Therefore the value of the equilibrium SFD relative to geometric saturation could vary significantly for impactor populations with different slope.

We also do not explicitly model secondary cratering in our CTEM simulations, even though primary mechanism by which distal degradation operates is through the formation of secondary craters. However, the explicit modeling of secondary craters is not likely to be necessary for the simulation of small crater equilibrium in the lunar maria. First, on the Moon, the largest secondary craters are typically no more than 4% the size of the primary, and the secondary SFD typically has a steeper slope than the primary production SFD (Melosh, 1989). For our 1 m/pix CTEM simulation, the smallest craters we can reliably model with enough fidelity to count are those with $r > 2.5$ m. Therefore the smallest craters that could produce countable secondaries are those with $r \sim 60$ m. In our simulations, there are usually no more than a few craters that form of this size or larger, and so including their secondaries would add insignificant numbers of countable craters to the surface.

The inclusion of the superdomain in our CTEM simulations allows for the distal effects of craters larger than those that are explicitly modeled on the domain to affect the simulation. Indeed, the inclusion of distal degradation by the superdomain craters is critical to our model. However, for these distant superdomain craters, explicit modeling of secondaries is still not likely to be necessary. Even though the largest secondaries are 4% of the primary crater size, the typical secondary populations found in distal features, such as rays, are far smaller. For instance, Fig. (2) of Elliott et al. (2018) shows a high resolution (50 cm/pix) image of a portion of the distal ejecta of the $r = 16$ km Kepler crater. This image shows that the energetic deposition of distal ejecta appears to have produced large numbers of small secondaries, o the order of 10s of m in size or less. Therefore, only the distal ejecta found in the rays of large complex craters produce distal secondaries in the size range of the craters of our study.

Our results are similar to the results of Hartmann and Gaskell (1997) who studied crater equilibrium on heavily cratered terrains of Mars. They showed that "sandblasting by subresolution secondary craters" was needed to match the equilibrium SFD on martian terrains based on results from cratering experiments using a three-dimensional Monte Carlo landscape evolution code that was very similar to our CTEM. However, Hartmann and Gaskell (1997) never quantified their sandblasting model. Conceptually, this degradation arises from the energetic deposition of distal ejecta, similar to ballistic sedimentation (Oberbeck, 1975). Our results provide new quantifiable constraints on the distal degradation that accompanies each new crater formation event.

Our model also demonstrates an important regarding the diffusive topographic evolution of the lunar surface that arises as a consequence of the observation that the equilibrium SFD of small craters has a slope of $\beta \sim 2$. An assumption adopted in many studies of the evolution of lunar landscapes is that the topographic diffusivity, $\kappa$ (which, under most circumstances is directly proportional to the degradation rate $K'$ used in our models) is the same at all size scales. For instance, Fassett and Thomson (2014) used craters in the range of $400\,\text{m} < r < 2500\,\text{m}$ to estimate an average diffusivity of the lunar surface of $\kappa \sim 5.5\,\text{m}^2/\text{My}$. However, this kind of constant and scale-independent degradation rate results in an equilibrium SFD with a slope that is $\beta \sim 1.2$, which significantly shallower than the observed value of $\beta \sim 2$. Therefore the degradation rate (or diffusivity) experienced by lunar craters must depend on crater size in such a way that small craters experience a degradation rate that is, on average over their lifetimes, *lower* than that of larger craters.

The requirement that the lunar surface experiences this type of scale dependence of absolute degradation rate was also suggested by Schultz et al. (1976), who noted that small scale topographic features associated with the emplacement of mare were apparently older than they should have been assuming a degradation lifetime constrained from the degradation of large craters. This result is somewhat counter-intuitive, because even though the slope of the equilibrium SFD requires that the diffusive degradation rate of smaller craters is lower than that of large craters, small craters require a lower value of accumulated degradation state in order to be fully obliterated compared with large craters. Therefore the lifetime of small craters will still be shorter than that of large craters, even if the degradation rate, $K'$ (or diffusivity, $\kappa$) experienced by smaller craters is lower than that experienced by larger craters.

## Acknowledgments

The authors acknowledge the support of NASA Lunar Data Analysis Program Grant no. NNX17AI79A. We thank reviewer Mikhail Kreslavsky, whose suggestions greatly improved our analytical models. We also thank reviewer Robert Craddock, whose comments lead to improvements in the structure and readability of the paper. We also thank Greg Michael, Clark Chapman, and H. Jay Melosh for their reviews of various earlier versions of this work. Finally, we would like to thank the editor at *Icarus*, Francis Nimmo, for both his thorough review and his exceptionally helpful editorial advice.